\documentclass[%
reprint,
superscriptaddress,
showpacs,preprintnumbers,
verbatim,
amsmath,amssymb,
aps,
prx,
]{revtex4-1}

\usepackage{color}
\usepackage{graphicx}% Include figure files
\usepackage{dcolumn}% Align table columns on decimal point
\usepackage{bm}% bold math
\usepackage{arydshln}% dashed line
\begin {document}
%section {title}
%\preprint{APS/123-QED}

\title{%
  Elucidating fluctuating diffusivity in center-of-mass motion of \\polymer
  models with time-averaged mean-square-displacement tensor}
% Force line breaks with \\
%\thanks{A footnote to the article title}%

%% Time-averaged mean-square-displacement tensor: a novel method to elucidate
%% fluctuating diffusivity

\author{Tomoshige Miyaguchi}
\email{tmiyaguchi@naruto-u.ac.jp}
\affiliation{%
  Department of Mathematics, 
  Naruto University of Education, Tokushima 772-8502, Japan}

%% \author{Takuma Akimoto}
%\email{akimoto@keio.jp}
%% \affiliation{%
%% Department of Mechanical Engineering, Keio University, Yokohama, 223-8522, Japan
%% }%

%%   \author{Eiji Yamamoto}
%%   \affiliation{%
%%   Department of Mechanical Engineering, Keio University, Yokohama, 223-8522, Japan
%% }%

%\collaboration{MUSO Collaboration}%\noaffiliation

\date{\today}% It is always \today, today,
%  but any date may be explicitly specified

\begin{abstract}
  There have been increasing reports that the diffusion coefficient of
  macromolecules depends on time and fluctuates randomly. Here a novel method is
  developed to elucidate this fluctuating diffusivity from trajectory data.  The
  time-averaged mean square displacement (MSD), a common tool in
  single-particle-tracking (SPT) experiments, is generalized to a second-order
  tensor, with which both magnitude and orientation fluctuations of the
  diffusivity can be clearly detected. This new method is used to analyze the
  center-of-mass motion of four fundamental polymer models: the Rouse model, the
  Zimm model, a reptation model, and a rigid rod-like polymer.  It is found that
  these models exhibit distinctly different types of magnitude and orientation
  fluctuations of the diffusivity. This is an advantage of the present method
  over previous ones such as the ergodicity-breaking parameter and a
  non-Gaussian parameter, because with either of these parameters it is
  difficult to distinguish the dynamics of the four polymer models. Also, the
  present method of a time-averaged MSD tensor could be used to analyze
  trajectory data obtained in SPT experiments.
\end{abstract}

%\pacs{05.45.Ac, 05.40.Fb, 87.15.Vv}% PACS, the Physics and Astronomy
% Classification Scheme.
%\keywords{Suggested keywords}%Use showkeys class option if keyword
%display desired
\maketitle

%\tableofcontents

\section {Introduction}
%subsection {general introduction}
%subsubsection {establish importance of this research topic}

Macromolecular diffusion in cytoplasm and cell membranes has received much
attention in recent years, because it controls chemical kinetics and information
processing in cells \cite{berg81}. Single-particle-tracking (SPT) techniques
have been used to study macromolecular diffusion in living systems, and
remarkably complicated phenomena such as anomalous diffusion, weak ergodicity
breaking (EB), and sample-to-sample fluctuations of the diffusion coefficient
have been reported \cite{golding06, weigel11, parry14, jeon11, tabei13}.
%% Also, much effort has been devoted to theoretical studies to try to describe
%% these complex behaviors by phenomenological models such as continuous-time
%% random walk (CTRW) and fractional Brownian motions \cite{he08, lubelski08,
%% deng09, neusius09, meroz10, miyaguchi11c, miyaguchi15a}.
%
%subsubsection {provide general background information}
%
In such SPT experiments, a time average is commonly used to obtain the mean
square displacement (MSD);
%% This is because relatively long trajectory data can be obtained in SPT
%% experiments.
the time-averaged MSD (TMSD) of a tagged particle is defined by \cite{
  nordlund1914, he08, metzler14}
\begin{equation}
  \label{e.tmsd-def}
  \overline{\delta \bm{r}^2}(\Delta, t)
  :=
  \frac {1}{t-\Delta}
  \int_{0}^{t-\Delta}
  \delta \bm{r}^2 (\Delta, t')\,dt',
\end{equation}
where $\Delta$ is a lag time and $t$ is the total measurement time. In addition,
a displacement vector $\delta \bm{r} (\Delta, t')$ is defined as $ \delta \bm{r}
(\Delta, t') := \bm{r} (t' + \Delta) - \bm{r} (t'),$ where $\bm{r}(t')$ is the
position vector of the tagged particle at time $t'$.  Thus, the TMSD
$\overline{\delta \bm{r}^2}(\Delta, t)$ can be obtained from a single trajectory
$\bm{r}(t')$.

%subsubsection {describe general problem or current research focus of the field}

In SPT experiments of macromolecules in living systems, sample-to-sample
fluctuations of the diffusion coefficient have been observed frequently
\cite{golding06, weigel11, parry14, jeon11, tabei13}. As stated above, the TMSD
curve $\overline{\delta \bm{r}^2}(\Delta, t)$ (as a function of $\Delta$) is
obtained from a single trajectory $\bm{r}(t')$, and then, from this TMSD curve,
the diffusion coefficient for that trajectory can be estimated.  The values of
this diffusion coefficient vary from trajectory to trajectory, but, for long
trajectories (namely, at $t\to \infty$), they converge to a single value if the
system is ergodic. In some SPT experiments, however, the values of the diffusion
coefficient are scattered even for long trajectories, and this phenomenon cannot
be explained by the ordinary Brownian motion \cite{golding06, jeon11,
  weigel11, tabei13, parry14}.

%subsection {literature review to solve the general problem}
%subsubsection {provide a brief overview of key research projections}

To explain such sample-to-sample fluctuation in the diffusivity, much effort has
been devoted to investigating simple theoretical models such as the
continuous-time random walk (CTRW) \cite{he08, lubelski08, neusius09,
  miyaguchi11c, miyaguchi13, thiel14b}, fractional Brownian motion \cite{deng09,
  thiel14}, and the random walk on fractals \cite{miyaguchi15a, meroz10}. In
these studies, the variance of the TMSD, which is commonly referred to as the EB
parameter, has been used to characterize the fluctuation in the diffusivity.
%% This parameter is related to the fourth order moment of the position vector
%% $\bm{r}(t)$.
In particular, it was shown that the EB parameter for the CTRW converges to a
non-vanishing value as $t\to \infty$. In other words, the TMSD behaves as a
random variable even for long measurement times. Therefore, CTRW-like dynamics
have been considered to be a factor in the sample-to-sample fluctuation of the
diffusivity observed in SPT experiments \cite{he08, lubelski08, neusius09}.

However, fluctuations in diffusivity originate also from correlated dynamics of
inner degrees of freedom. In Ref.~\cite{uneyama15}, the authors studied a
reptation model (a tagged polymer model in entangled polymer solutions) and
showed that the EB parameter of the center-of-mass (COM) motion is non-vanishing
for quite a long measurement time. In other words, the system exhibits
sample-to-sample fluctuations in diffusivity, that originate from non-Markovian
dynamics of the end-to-end vector.
%, and thus .
%% In addition, the authors show that the COM motions of the reptation model can
%% be described by a Langevin equation with time-dependent and fluctuating
%% diffusivity. One of the important findings of Ref.\cite{uneyama15} is that
%% the EB parameter is related to a magnitude correlation of the fluctuating
%% diffusivity.
%subsubsection {describe a gap in the research}
Another important finding of Ref.~\cite{uneyama15} is that the EB parameter is
related to a correlation function of magnitude of diffusivity.  Unfortunately,
it was also found that much of the information contained in the trajectory data
$\bm{r}(t)$ is lost in the EB parameter. Therefore, it is necessary to develop
an efficient method to extract more information from the trajectory data.
%% Little has been known whether the fluctuating diffusivity is general in the
%% COM motions of polymer dynamics other than the reptation model.

%subsection {describe the present paper}
%subsubsection {describe the present paper}
In this paper, a novel method is developed for elucidating the fluctuating
diffusivity of macromolecules from trajectory data $\bm{r}(t)$.
%subsubsection {describe the methodology reported in the present paper}
%subsubsection {announce the findings}
More precisely, a TMSD tensor, a generalization of the TMSD
[Eq.~(\ref{e.tmsd-def})], is proposed, and it is shown that correlation
functions of this TMSD tensor contain plenty of information including a
magnitude correlation and an orientation correlation of the fluctuating
diffusivity. Moreover, by using this tensor analysis, four fundamental polymer
models are investigated: the Rouse and Zimm models (polymer models in dilute
solutions), a reptation model (a polymer model in concentrated solutions), and a
rigid rod-like polymer (an extreme case of non-flexible polymers). It is shown
that the COM motion of these polymer models exhibits distinctly different types
of the fluctuating diffusivity. For example, it is shown that the COM motion of
the Zimm and reptation models exhibits both magnitude and orientation
fluctuations of the diffusivity, whereas that of the rigid rod-like polymer
exhibits only orientation fluctuations. The tensor analysis presented in this
article could be used to analyze the trajectory data obtained in SPT
experiments.
%
%
%% What is the physical meaning of the ergodicity breaking parameter?  shown
%% that the EB parameter is related to the magnitude correlation of the
%% diffusion coefficient. In addition,
%

%% By using this method, we study the three polymer models above-mentioned. 

%subsubsection {organization of the paper}

This paper is organized as follows. In Sec.~\ref{s.model}, a Langevin equation
with fluctuating diffusivity (LEFD) is defined. In Sec.~\ref{s.def.tmsd}, the
TMSD tensor is defined and its correlation functions are studied for the LEFD.
It is also shown here that these correlation functions are related to a
non-Gaussian parameter. In Secs.~\ref{s.rouse_model}--~\ref{s.rodlike_polymer},
the COM motion of each of the aforementioned polymer models is studied with the
TMSD tensor. Finally, Sec.~\ref{s.discussion} is devoted to a discussion. In the
Appendices, we summarize some technical matters, including the simulation
details.

\section {Langevin equation with fluctuating Diffusivity}\label{s.model}

As shown in subsequent sections, the COM of polymer models such as the Zimm and
reptation models can be described by the following Langevin equation with
time-dependent and fluctuating diffusivity \cite{stein91, luczka95, rozenfeld98,
  luczka00, mykyta14, massignan14, manzo15, cherstvy16, uneyama15, miyaguchi16,
  chechkin17}:
\begin{equation}
  \label{e.def-lefd}
  \frac {d\bm{r}(t)}{dt} = \sqrt{2}\,\bm{B}(t) \cdot \bm{\xi}(t),
\end{equation}
where $\bm{r}(t)$ is an $n$-dimensional position vector of a tagged particle at
time $t$, and the $n\times n$ matrix $\bm{B}(t)$ is a stochastic
process. Moreover, $\bm{\xi}(t)$ is white Gaussian noise that satisfies
\begin{equation}
  \label{e.def-lefd.<xi(t)xi(0)>}
  \left\langle \bm{\xi}(t)\bm{\xi}(t') \right\rangle
  =
  \bm{I} \delta(t-t'),
\end{equation}
where $\bm{I}$ is the identity matrix.
%
%% Stochastic processes similar to Eq.~(\ref{e.def-lefd}) have been widely studied
%% in various contexts \cite{stein91, luczka95, rozenfeld98, luczka00, mykyta14,
%%   massignan14, manzo15, cherstvy16, miyaguchi16, chechkin17}.
%% In this paper,
Equation (\ref{e.def-lefd}) is referred to as the LEFD.

In this study, it is assumed that $\bm{\xi}(t)$ and $\bm{B}(t)$ are mutually
independent stochastic processes. Consequently, the diffusion coefficient tensor
$\bm {D}(t)$ is given by
\begin{equation}
  \label{e.D(t)=B(t).B^T(t)}
  \bm{D}(t) = \bm{B}(t)\cdot\bm{B}^T(t),
\end{equation}
where $\bm{B}^T$ is the transpose matrix of $\bm{B}$. It follows that
$\bm{D}(t)$ is a symmetric tensor: $\bm{D}(t) = \bm{D}^T(t)$.  In addition,
$\bm{D}(t)$ is assumed to be a stationary process.

%% the system is assumed to be in equilibrium in what follows. In other words, the
%% fluctuating diffusivity $\bm{D}(t)$ arises from an equilibrium process.
%

%heston93, scott97,

\section {TMSD tensor}\label{s.def.tmsd}
%subsection {intro}

In this section, the TMSD tensor is defined and its general properties are
presented. In particular, it is shown that the TMSD tensor of the LEFD exhibits
only normal diffusion, even though the density profile is
non-Gaussian. Moreover, to extract information on the fluctuating diffusivity,
correlation functions of the TMSD tensor are studied. In particular, a novel
method to extract magnitude and orientation correlations of the diffusivity is
presented.

\subsection {TMSD tensor exhibits normal diffusion}

As a generalization of the TMSD [Eq.~(\ref{e.tmsd-def})], a TMSD tensor (a
second-order tensor) is defined as
\begin{equation}
  \label{e.tmsd-tensor}
  \overline{\delta \bm{r}\delta \bm{r}}(\Delta, t)
  :=
  \frac {1}{t-\Delta}
  \int_{0}^{t-\Delta}
  \delta \bm{r} (\Delta, t')
  \delta \bm{r} (\Delta, t')\,dt',
\end{equation}
where the integral is taken for each element of the tensor in the
integrand as
\begin{equation}
  \label{e.tmsd-tensor.component}
  \left[\overline{\delta \bm{r}\delta \bm{r}}(\Delta, t)\right]_{ij}
  =
  \frac {1}{t-\Delta}
  \int_{0}^{t-\Delta}
  \delta {r}_{i} (\Delta, t')
  \delta {r}_{j} (\Delta, t')\,dt'.
\end{equation}
Here $\delta r_{i} (\Delta, t')$ is an element of $\delta \bm{r} (\Delta, t')$,
and $\left[ \bm{H} \right]_{ij}$ represents an element of a second-order tensor
$\bm{H}$: $\left[ \bm{H} \right]_{ij}:=H_{ij}$.

Note that the TMSD tensor $\overline{\delta \bm{r}\delta \bm{r}}(\Delta, t)$ is
the time-averaged counterpart of the ensemble-averaged MSD tensor
\cite{dhont96}.
%% The diagonal elements of the TMSD tensor are time-averaged square
%% displacements in $x, y$ and $z$ directorsこの TMSD テンソルの対角成分はそれぞ
%% れ, $x,y,z$ 方向への二乗変位の時間平均を表す. 一方, 非対角成分は異る方向への変
%% 位間の (時間平均による) 相関を意味する.
Taking the trace of Eq.~(\ref{e.tmsd-tensor}), we obtain the TMSD given in
Eq.~(\ref{e.tmsd-def}), and thus it is possible to extract more information with
the TMSD tensor than with the TMSD. Moreover, taking the ensemble average in
Eq.~(\ref{e.tmsd-tensor}) and using
Eqs.~(\ref{e.def-lefd})--(\ref{e.D(t)=B(t).B^T(t)}), we have
\begin{align}
  \left\langle
  \overline{\delta\bm{r}\delta\bm{r}}(\Delta, t)
  \right\rangle
  &=
  \left\langle
  \delta \bm{r} (\Delta, 0)
  \delta \bm{r} (\Delta, 0)  
   \right\rangle
  \notag\\[0.1cm]
  &=
  2\!
  \int_{0}^{\Delta} \!\!dt_1 \!
  \int_{0}^{\Delta} \!\!dt_2
  \left\langle
  \bm{B}(t_1)\cdot \bm{\xi}(t_1)
  \bm{B}(t_2)\cdot \bm{\xi}(t_2)
  \right\rangle
  \notag\\[0.1cm]
  \label{e.et_msd_tensor}
  &= 2 \left\langle \bm{D} \right\rangle \Delta,
\end{align}
where $\left\langle\dots\right\rangle$ is the ensemble average. For the first
equality in Eq.~(\ref{e.et_msd_tensor}), we used the stationarity of the system,
and for the final equality, we used the fact that $\bm{B}(t)$ and
$\bm{\xi}(t)$ are independent in the sense that
\begin{align}
  \left\langle
  B_{ik}(t_1)\xi_{k}(t_1)
  B_{jl}(t_2)\xi_{l}(t_2)
  \right\rangle  &
  \notag\\[0.1cm]
  %% &=
  %% \left\langle
  %% B_{\alpha\gamma}(t_1)B_{\beta\delta}(t_2)
  %% \right\rangle
  %% \left\langle
  %% \xi_{\gamma}(t_1)\xi_{\delta}(t_2)
  %% \right\rangle
  %% \notag\\[0.1cm]
  & \hspace*{-2.5cm}=
  \left\langle
  B_{ik}(t_1)B_{jl}(t_2)
  \right\rangle
  \delta_{kl}\delta(t_1 -t_2),
\end{align}
where we have employed the Einstein summation convention. In particular, if the
system is statistically isotropic,
%% (i.e., the ensemble averages of the tensors are independent of the choice of the
%% orthogonal basis)
we have $\left\langle \bm{D} \right\rangle = D\, \bm{I}$. Taking the trace in
Eq.~(\ref{e.et_msd_tensor}), we obtain the TMSD again \cite{uneyama15}
\begin{equation}
  \label{e.et_msd}
  \left\langle
  \overline{\delta\bm{r}^2}(\Delta, t)
  \right\rangle
  = 2 \,\mathrm{tr}\left\langle \bm{D} \right\rangle \Delta.
\end{equation}
Surprisingly, all the elements of the ensemble-averaged TMSD tensor in
Eq.~(\ref{e.et_msd_tensor}) exhibit only normal diffusion (i.e., proportional to
the lag time $\Delta$), even though the diffusion coefficient fluctuates. In
other words, it is impossible to detect the fluctuating diffusivity with the
first moment of the TMSD tensor [Eq.~(\ref{e.et_msd_tensor})], and so
higher-order moments of the TMSD tensor are studied in the following
subsections.

\subsection {Correlation function of TMSD tensor}

To extract information about the fluctuating diffusivity from trajectories
$\bm{r}(t)$, we study a correlation function of the TMSD tensor
\begin{widetext}
  \begin{align}
    \label{e.phi-4th-tensor.def}
    \bm{\Phi}(\Delta, t)
    :=&
    \left\langle 
    \left[\,
    \overline{\delta \bm{r}\delta \bm{r}}(\Delta, t)
    -
    \left\langle 
    \overline{\delta \bm{r}\delta \bm{r}}(\Delta, t)
    \right\rangle
    \right]
    \left[\,
    \overline{\delta \bm{r}\delta \bm{r}}(\Delta, t)
    -
    \left\langle 
    \overline{\delta \bm{r}\delta \bm{r}}(\Delta, t)
    \right\rangle
    \right]
    \right\rangle
    \\[0.1cm]
    \label{e.phi-4th-tensor}
    =&
    \left\langle 
    \overline{\delta \bm{r}\delta \bm{r}}(\Delta, t)\,
    \overline{\delta \bm{r}\delta \bm{r}}(\Delta, t)
    \right\rangle
    -
    \left\langle 
    \overline{\delta \bm{r}\delta \bm{r}}(\Delta, t)\,
    \right\rangle
    \left\langle 
    \overline{\delta \bm{r}\delta \bm{r}}(\Delta, t)
    \right\rangle,
  \end{align}
\end{widetext}
where $\bm{\Phi}(\Delta, t)$ is a fourth-order tensor. Note that, in time-series
analysis, Eq.~(\ref{e.phi-4th-tensor.def}) should be used instead of
Eq.~(\ref{e.phi-4th-tensor}) to reduce numerical errors. In fact,
Eq.~(\ref{e.phi-4th-tensor.def}) was used in all of the numerical simulations
reported here.

If we assume that $\Delta$ is much shorter than a characteristic time scale
$\tau_D$ of the fluctuating diffusivity, we can decompose $\bm{\Phi}(\Delta,
t)$ into two parts (see below for a derivation) as
%% ここで, 式(\ref{e.tmsd-tensor}) を代入して, $\Delta$ が拡散係数揺らぎの特徴的
%% タイムスケールより十分小さいと仮定すれば,
\begin{equation}
  \label{e.phi-4th-tensor.phi_id+phi_ex}
  \bm{\Phi}(\Delta, t)
  \approx
  \bm{\Phi}^{\mathrm{id}} (\Delta, t)
  +
  \bm{\Phi}^{\mathrm{ex}} (\Delta, t),
\end{equation}
where the fourth-order tensors $\bm{\Phi}^{\mathrm{id}} (\Delta, t)$ and
$\bm{\Phi}^{\mathrm{ex}} (\Delta, t)$ are defined respectively as
\begin{align}
  \label{e.phi-4th-tensor.phi_id}
  \Phi_{ikmp}^{\mathrm{id}} (\Delta, t)
  =
  \frac {2\Delta^3}{3t}
  \left( 4 - \frac {\Delta}{t}\right)
  \left(
  \left\langle D_{im}D_{kp} \right\rangle
  +
  \left\langle D_{ip}D_{km} \right\rangle
  \right),
  \\[0.1cm]
  \label{e.phi-4th-tensor.phi_ex}
  \bm{\Phi}^{\mathrm{ex}}(\Delta, t)
  =
  \frac {8\Delta^2}{t^2}\!
  \int_{0}^{t} \!d\tau (t-\tau)
  \Bigl[
  \left\langle \bm{D}(\tau) \bm{D}(0) \right\rangle_s
  -
  \left\langle \bm{D} \right\rangle\!
  \left\langle \bm{D} \right\rangle\Bigr].
\end{align}
Here, $\left\langle \dots \right\rangle_s$ is a symmetrization given by
\begin{equation}
  \label{e.<...>_s}
  \left\langle \bm{D}(\tau) \bm{D}(0) \right\rangle_s
  :=
  \frac
  {\left\langle \bm{D}(\tau) \bm{D}(0) \right\rangle
    +
    \left\langle \bm{D}(0) \bm{D}(\tau) \right\rangle}
  {2}.
\end{equation}

Equation (\ref{e.phi-4th-tensor.phi_id+phi_ex}) can be derived as follows.
First, $\bm{\Phi}(\Delta, t)$ is expressed as $\bm{\Phi}(\Delta, t) =
\bm{\Psi}^1(\Delta, t) - \bm{\Psi}^2(\Delta, t)$, where $\bm{\Psi}^1(\Delta, t)$
and $\bm{\Psi}^2(\Delta, t)$ are fourth-order tensors defined [see
Eq.~(\ref{e.phi-4th-tensor})] as
\begin{align}
  \label{e.def.Psi1}
  \bm{\Psi}^1(\Delta, t)
  &:=
  \left\langle
  \overline{\delta \bm{r}\delta\bm{r}}(\Delta, t)\,
  \overline{\delta \bm{r}\delta\bm{r}}(\Delta, t)
  \right\rangle,
  \\[0.1cm]
  \label{e.def.Psi2}
  \bm{\Psi}^2(\Delta, t)
  &:=
  \left\langle
  \overline{\delta \bm{r}\delta \bm{r}}(\Delta, t)
  \right\rangle
  \left\langle
  \overline{\delta \bm{r}\delta\bm{r}}(\Delta, t)
  \right\rangle.
\end{align}
After a lengthy calculation, the elements of $\bm{\Psi}^1(\Delta, t)$ can be
expressed (see Appendix \ref{s.derivation.Phi=Phi_id+Phi_ex} for detail) as
\begin{align}
  \label{e.final.Psi1}
  \Psi_{ikmp}^1 (\Delta, t)
  &=
  \Phi_{ikmp}^{\mathrm{id}}(\Delta, t)
  \notag\\[0.1cm]
  +
  &
  \frac {8\Delta^2}{t^2}
  \int_0^t ds (t-s)
  \Bigl[
  \left\langle \bm{D}(s) \bm{D}(0) \right\rangle_s
  \Bigr]_{ikmp},
  %\left\langle D_{ik}(s) D_{mp}(0) \right\rangle
\end{align}
where $\Phi_{ikmp}^{\mathrm{id}}(\Delta, t)$ is the ideal part defined in
Eq.~(\ref{e.phi-4th-tensor.phi_id}), and $\left[ \bm{H} \right]_{ikmp}$
represents an element of a fourth-order tensor $\bm{H}$, i.e., $\left[ \bm{H}
\right]_{ikmp}:=H_{ikmp}$.  On the other hand, from Eqs.~(\ref{e.et_msd_tensor})
and (\ref{e.def.Psi2}), we have
\begin{align}
  \bm{\Psi}^2(\Delta, t)
  &=
  4 \Delta^2
  \left\langle \bm{D} \right\rangle\!
  \left\langle \bm{D} \right\rangle
  \label{e.Psi2}
  =
  \frac {8\Delta^2}{t^2}
  \int_0^t ds (t-s)
  \left\langle \bm{D} \right\rangle\!
  \left\langle \bm{D} \right\rangle.
\end{align}
By subtracting Eq.~(\ref{e.Psi2}) from Eq.~(\ref{e.final.Psi1}), the elements of
the fourth-order tensor $\bm{\Phi}(\Delta, t)$ are obtained as
\begin{align}
  \label{e.app.Phi=Phi^id+Phi^ex}
  \Phi_{ikmp}(\Delta, t)
  & =
  \Phi_{ikmp}^{\mathrm{id}}(\Delta, t)
  \notag\\[0.1cm]
  +
  \frac {8\Delta^2}{t^2}\!\!
  &
  \int_0^t ds (t-s)\!
  \Bigl[
  \left\langle \bm{D}(s) \bm{D}(0) \right\rangle_s
  -
  \left\langle \bm{D} \right\rangle\!
  \left\langle \bm{D} \right\rangle
  \Bigr]_{ikmp}.
\end{align}
The second term in the right-hand side is equivalent to
$\Phi_{ikmp}^{\mathrm{ex}}(\Delta, t)$ [see
Eq.~(\ref{e.phi-4th-tensor.phi_ex})], and hence
Eq.~(\ref{e.app.Phi=Phi^id+Phi^ex}) coincides with
Eq.~(\ref{e.phi-4th-tensor.phi_id+phi_ex}).

As can be seen from Eq.~(\ref{e.phi-4th-tensor.phi_ex}), the tensor
$\bm{\Phi}^{\mathrm{ex}}(\Delta, t)$ is related to the autocorrelation function
of the diffusivity tensor $\bm{D}(t)$. Thus, in contrast to the first moment of
the TMSD tensor given in Eq.~(\ref{e.et_msd_tensor}), the second moment
$\bm{\Phi}(\Delta, t)$ can be used to characterize the fluctuating
diffusivity. In particular, if $\bm{D}(t)$ does not fluctuate, then
$\bm{\Phi}^{\mathrm{ex}}(\Delta, t) \equiv 0$; therefore,
$\bm{\Phi}^{\mathrm{ex}}(\Delta, t)$ is hereinafter referred to as an excess
part. In contrast, the qualitative features of $\bm{\Phi}^{\mathrm{id}} (\Delta,
t)$ in Eq.~(\ref{e.phi-4th-tensor.phi_id}) are independent of the fluctuating
diffusivity,
%(a difference arises only from the constant factor
% $\left\langle D_{im}D_{kp} \right\rangle + \left\langle D_{ip}D_{km} \right\rangle$),
and therefore this part is referred to as an ideal part.

%% An important point is that, in single-particle-tracking experiments, the
%% instantaneous diffusion coefficient $\bm{D}(t)$ is quite difficult to measure
%% directly. This is because a particle displacement is related not only to
%% $\bm{D}(t)$ but also to the thermal noise $\bm{\xi}(t)$ [see
%% Eq.~(\ref{e.def-lefd})]. For example, a large displacement does not
%% necessarily mean a large value of $\bm{D}(t)$, since it may be $\bm{\xi}(t)$
%% that takes a large value. However,
An important point is that the TMSD tensor $\overline{\delta \bm{r}\delta
  \bm{r}}(\Delta, t) $ and its correlation function $\bm{\Phi}(\Delta, t)$ can
be calculated from the trajectory data $\bm{r}(t)$ alone, and there is no need
to measure $\bm{D}(t)$. Since the trajectory data $\bm{r}(t)$ is available in
many single-particle-tracking experiments, the TMSD tensor and its correlation
function are useful tools for elucidating the fluctuating diffusivity. Note
however that in the derivation of Eq.~(\ref{e.phi-4th-tensor.phi_id+phi_ex}), it
is assumed that $\Delta$ is shorter than a characteristic time scale $\tau_D$ of
the fluctuating diffusivity. This means that the observation interval should be
much shorter than $\tau_D$.

%% したがって, 4 階テンソル $\bm{\Phi}^{\mathrm{ex}}(\Delta, t)$ は拡散係数
%% の相関関数の情報を有していることが分かる. ここで注意したい点は, 瞬間の拡散係
%% 数 $\bm{D}(t)$ は直接測定することが難しいのに対し, $\bm{\Phi}(\Delta, t)$ は軌道
%% の時系列データ$\bm{r}(t)$ だけから計算でき, 瞬間の拡散係数 $\bm{D}(t)$ を測定する
%% 必要がない,ということである (ただし, 上式を導出する際に $t/\Delta \gg 1$ という条
%% 件を使用している. つまり, $\Delta$ をできるだけ小さく取る必要があるが, そのために
%% は軌道時系列データの時間解像度をできるだけ細かくする必要がある).

\subsection {Correlation functions of diffusion coefficient}

To obtain more specific information of the fluctuating diffusivity, two scalar
functions $\Phi_1(\Delta, t)$ and $\Phi_2(\Delta, t)$ are derived from
$\bm{\Phi}(\Delta, t)$. It is shown that these are related to a magnitude and
orientation correlations, respectively, of the fluctuating diffusivity
$\bm{D}(t)$.

\subsubsection {Magnitude correlation of diffusion coefficient}

Firstly, $\Phi_1 (\Delta, t)$ is defined as a scalar quantity obtained by taking
contractions in Eqs.~(\ref{e.phi-4th-tensor}) or
(\ref{e.phi-4th-tensor.phi_id+phi_ex}) between the first and second indices, and
also between the third and fourth indices. It follows that $\Phi_1 (\Delta,
t)$ is given by
\begin{align}
  \label{e.Phi_1=Phi_1^id+Phi_1^ex}
  \Phi_1(\Delta, t)
  =&
  \left\langle 
  |\overline{\delta \bm{r}^2}(\Delta, t)|^2
  \right\rangle
  -
  \left\langle 
  \overline{\delta \bm{r}^2}(\Delta, t)
  \right\rangle^2 \notag\\[0.1cm]
  \approx&
  \Phi_1^{\mathrm{id}}(\Delta, t) +
  \Phi_1^{\mathrm{ex}}(\Delta, t),
\end{align}
where the two scalar functions $\Phi_1^{\mathrm{id}}(\Delta, t)$ and
$\Phi_1^{\mathrm{ex}}(\Delta, t)$ are defined by
\begin{align}
  \Phi_1^{\mathrm{id}}(\Delta, t) :=&
  \frac {4\Delta^3}{3t}
  \left( 4 - \frac {\Delta}{t}\right)
  \mathrm{tr} \left\langle \bm{D}\cdot \bm{D} \right\rangle,
  \\[0.1cm]
  \Phi_1^{\mathrm{ex}}(\Delta, t) :=&
  \frac {8\Delta^2}{t^2}\! \int_{0}^{t} \! d\tau (t-\tau) 
  %% \\[0.1cm]\notag
  %% &
  %% \hspace*{.5cm}\times
  \left[
  \left\langle \mathrm{tr}\bm{D}(\tau) \mathrm{tr}\bm{D}(0) \right\rangle
  -
  \langle \mathrm{tr} \bm{D}\rangle^2
  \right].
\end{align}
As can be seen from Eq.~(\ref{e.Phi_1=Phi_1^id+Phi_1^ex}), $\Phi_1(\Delta, t)$
is the variance of the TMSD [Eq.~(\ref{e.tmsd-def})].

Furthermore, Eq.~(\ref{e.Phi_1=Phi_1^id+Phi_1^ex}) can be made dimensionless by
dividing it by $ \langle \overline{\delta \bm{r}^2}(\Delta, t) \rangle^2 = 4
\Delta^2 \langle \mathrm{tr} \bm{D}\rangle^2$; this dimensionless quantity is
denoted as $\hat{\Phi}_1(\Delta, t)$ and is given by
\begin{equation}
  \label{e.hatPhi_1=hatPhi_1^id+hatPhi_1^ex}
  \hat{\Phi}_1(\Delta, t)\approx
  \hat{\Phi}_1^{\mathrm{id}}(\Delta, t) + \hat{\Phi}_1^{\mathrm{ex}}(t).
\end{equation}
Note that $\hat{\Phi}_1(\Delta, t)$ is the relative variance of the TMSD, which
is equivalent to the EB parameter \cite{he08, deng09, uneyama15, miyaguchi16}.
The two scalar functions $\hat{\Phi}_1^{\mathrm{id}}(\Delta, t)$ and
$\hat{\Phi}_1^{\mathrm{ex}}(t)$ are defined respectively as
\begin{align}
  \label{e.hatPhi1_id}
  \hat{\Phi}_1^{\mathrm{id}}(\Delta, t)
  &:=
  \frac {C\Delta}{3nt} \left(4 - \frac {\Delta}{t}\right)
  \left[1 + \phi_2(0)\right],
  \\[0.1cm]
  \label{e.hatPhi1_ex}
  \hat{\Phi}_1^{\mathrm{ex}}(t)
  &:=
  \frac {2}{t^2}
  \int_{0}^{t} d\tau (t-\tau)
  \phi_1(\tau).
\end{align}
Here, $n$ is the space dimension, $\phi_1(\tau)$ and $\phi_2(\tau)$ are
magnitude and orientation correlation functions, respectively, of the
diffusivity $\bm{D}(t)$:
\begin{align}
  \label{e.phi1}
  \phi_1(\tau)
  &:=
  \frac
  {\left\langle \mathrm{tr}\bm{D}(\tau)\, \mathrm{tr}\bm{D}(0)\right\rangle}
  {\left(\mathrm{tr} \langle\bm{D}\rangle\right)^2}
  -1,
  \\[0.2cm]
  \label{e.phi2}
  \phi_2(\tau)
  &:=
  \frac
  {\mathrm{tr}\left\langle \bm{D}(\tau) \cdot \bm{D}(0) \right\rangle}
  {\mathrm{tr} \left(\langle \bm{D}\rangle\cdot\langle\bm{D}\rangle\right)}
  -1,
\end{align}
and $C$ is a constant defined by
\begin{equation}
  C
  := n
  \frac
  {\mathrm{tr} \left(
    \left\langle \bm{D} \right\rangle \cdot
    \left\langle \bm{D} \right\rangle
    \right)}
  {\left(\mathrm{tr} \left\langle \bm{D} \right\rangle\right)^{2}}.
\end{equation}
If the system is statistically isotropic, then we have $\left\langle \bm{D}
\right\rangle = D \bm{I}$ and hence $C= 1$.

As seen from Eq.~(\ref{e.hatPhi1_ex}), $\hat{\Phi}_1^{\mathrm{ex}}(\Delta, t)$
is related to the magnitude correlation function $\phi_1(\tau)$ of the
diffusivity. For example, if the magnitude of the diffusivity is constant [i.e.,
$\mathrm{tr}\bm{D}(t) \equiv \mathrm{const.}$] and only its direction
fluctuates, we have $\hat{\Phi}_1^{\mathrm{ex}}(\Delta, t) \equiv 0$ from
Eqs.~(\ref{e.hatPhi1_ex}) and (\ref{e.phi1}); thus, no information about the
fluctuating diffusivity can be detected with $\hat{\Phi}_1(\Delta, t)$. This is
actually the case for the COM motion of the rigid rod-like polymer
(Sec.~\ref{s.rodlike_polymer}), and it is necessary to study a different
quantity to elucidate the orientation fluctuation.

\subsubsection {Orientation correlation of diffusion coefficient}

To extract information about the orientation fluctuation, another scalar
function $\Phi_2 (\Delta, t)$ is defined by taking contractions in
Eqs.~(\ref{e.phi-4th-tensor}) or (\ref{e.phi-4th-tensor.phi_id+phi_ex}) both
between the second and third indices, and also between the first and fourth
indices. Consequently, $\Phi_2 (\Delta, t)$ is given by
\begin{align}
  \Phi_2(\Delta, t)
  =&
  \left\langle 
  \overline{\delta \bm{r}\delta \bm{r}}(\Delta, t):
  \overline{\delta \bm{r}\delta \bm{r}}(\Delta, t)
  \right\rangle
  \notag\\[0.1cm]
  &
  -
  \left\langle 
  \overline{\delta \bm{r}\delta \bm{r}}(\Delta, t)\,
  \right\rangle
  :
  \left\langle 
  \overline{\delta \bm{r}\delta \bm{r}}(\Delta, t)
  \right\rangle
  \\[.1cm]
  \label{e.Phi2}
  \approx&
  \Phi_2^{\mathrm{id}}(\Delta, t) +
  \Phi_2^{\mathrm{ex}}(\Delta, t),
\end{align}
where a double dot product "$:$" is defined by $\bm{A}\!:\!\bm{B} = \sum_{ij}^{}
A_{ij}B_{ji}$, and $\Phi_2^{\mathrm{id}}(\Delta, t)$ and
$\Phi_2^{\mathrm{ex}}(\Delta, t)$ are scalar functions defined respectively as
\begin{align}
  \Phi_2^{\mathrm{id}}(\Delta, t)
  =&
  \frac {2\Delta^{3}}{3t}
  \left(4 - \frac {\Delta}{t}\right)
  \left[
  \mathrm{tr} \left\langle \bm{D}\cdot \bm{D} \right\rangle
  +
  \left\langle (\mathrm{tr} \bm{D})^2 \right\rangle
  \right],
  \\[0.1cm]
  \Phi_2^{\mathrm{ex}}(\Delta, t)
  =&
  \frac {8\Delta^2}{t^2} \int_{0}^{t}  d\tau (t-\tau)
  \notag\\[0.1cm]
  &
  \hspace*{0.1cm}\times
  \left[
  \mathrm{tr}\left\langle \bm{D}(\tau) \cdot \bm{D}(0) \right\rangle
  -
  \mathrm{tr} \left(\langle \bm{D}\rangle\cdot\langle \bm{D}\rangle\right)
  \right].
\end{align}

Again, let us make Eq.~(\ref{e.Phi2}) dimensionless by dividing it by
$
\langle 
\overline{\delta \bm{r}\delta \bm{r}}(\Delta, t)\,
\rangle
:
\left\langle 
\overline{\delta \bm{r}\delta \bm{r}}(\Delta, t)
\right\rangle
=
4 \Delta^2 \mathrm{tr} \left(\langle \bm{D}\rangle\cdot\langle
\bm{D}\rangle\right)$;
we denote this dimensionless quantity as
$\hat{\Phi}_2(\Delta, t)$, which is given by
\begin{equation}
  \label{e.hatPhi_2=hatPhi_2^id+hatPhi_2^ex}
  \hat{\Phi}_2(\Delta, t)
  \approx
  \hat{\Phi}_2^{\mathrm{id}}(\Delta, t) + \hat{\Phi}_2^{\mathrm{ex}}(t),
\end{equation}
where the two scalar functions $\hat{\Phi}_2^{\mathrm{id}}(\Delta, t)$ and
$\hat{\Phi}_2^{\mathrm{ex}}(\Delta, t)$ are defined as
\begin{align}
  \label{e.hatPhi2_id}
  \hat{\Phi}_2^{\mathrm{id}}(\Delta, t)
  &=
  \frac {\Delta}{6t} \left(4 - \frac {\Delta}{t}\right)
  \left\{
  \frac {n}{C} [\phi_1(0) + 1] + \phi_2(0) + 1
  \right\},
  \\[0.1cm]
  \label{e.hatPhi2_ex}
  \hat{\Phi}_2^{\mathrm{ex}}(t)
  &=
  \frac {2}{t^2}
  \int_{0}^{t} d\tau (t-\tau)
  \phi_2(\tau).
\end{align}
The function $\phi_2(\tau)$, which is defined in Eq.~(\ref{e.phi2}), represents
an orientation correlation of the diffusivity, and hence information about the
orientation correlation can be extracted by using $\hat{\Phi}_2(\Delta, t)$.
Note however that, for the case in which the diffusivity tensor $\bm{D}(t)$ is
given by a scalar function $D(t)$ as $\bm{D}(t) = D(t) \bm{I}$, the two
functions $\hat{\Phi}_1^{\mathrm{ex}}(t)$ and $\hat{\Phi}_2^{\mathrm{ex}}(t)$
are equivalent: $\hat{\Phi}_1^{\mathrm{ex}}(t) = \hat{\Phi}_2^{\mathrm{ex}}(t)$.
%% This is because, from Eqs.~(\ref{e.phi1}) and (\ref{e.phi2}), we have
%% $\phi_1(\tau) = \phi_2(\tau)$ if $\bm{D}(t) = D(t) \bm{I}$.
In this sense, $\phi_2(\tau)$ includes information about the magnitude
correlation of the diffusivity as well as its orientation correlation; therefore
$\tilde{\phi}_2(\tau) := \phi_2(\tau) - \phi_1(\tau)/C$ may be more suitable as
an orientation correlation.
In what follows, however, $\phi_2(\tau)$ and $\hat{\Phi}_2^{\mathrm{ex}}(t)$ are
referred to as orientation correlation functions for simplicity.
%% In addition, for the case in which $\bm{D}(t) = D(t) \bm{I}$, it is enough to
%% observe one of the two sclar functions, $\hat{\Phi}_1^{\mathrm{ex}}(t)$ or
%% $\hat{\Phi}_2^{\mathrm{ex}}(t)$.
The special case in which $\bm{D}(t) = D(t) \bm{I}$ was studied extensively in
Ref.\cite{miyaguchi16}.

\subsection {Non-Gaussian parameter}

A non-Gaussian parameter of the displacement vector $\delta \bm{r}(t) =
\bm{r}(t) - \bm{r}(0)$ is defined as \cite{kegel00, arbe02, ernst14, cherstvy14}
\begin{align}
  A(t) := \frac {n}{n+2}
  \frac
  {\left\langle \delta \bm{r}^4(t) \right\rangle}
  {\left\langle \delta \bm{r}^2(t) \right\rangle^2} - 1.
\end{align}
In Ref.~\cite{uneyama15}, it was shown that the non-Gaussian parameter $A(t)$
for the LEFD [Eq.~(\ref{e.def-lefd})] is given by
\begin{equation}
  \label{e.non-gaussian-parameter}
  A(t)
  =
  \frac {2(C-1)}{n+2}
  +
  \frac {n}{n+2}
  \left[
  \hat{\Phi}_1^{\mathrm{ex}}(t)
  +
  \dfrac {2C}{n}\hat{\Phi}_2^{\mathrm{ex}}(t)
  \right].
\end{equation}
For isotropic systems, we have $C=1$; and hence the first term vanishes.
Equation (\ref{e.non-gaussian-parameter}) shows that the non-Gaussian parameter
$A(t)$ can be decomposed into two parts; one originates from the magnitude
correlation of the diffusivity, and the other from its orientation
correlation. Although Eq.~(\ref{e.non-gaussian-parameter}) was derived
previously in Ref.~\cite{uneyama15}, it was not known then how to calculate
$\hat{\Phi}_2^{\mathrm{ex}}(t)$ from the trajectory data $\bm{r}(t)$. Therefore,
the method for obtaining $\hat{\Phi}_2^{\mathrm{ex}}(t)$ as presented in the
previous subsection is one of the main results of this article.

\subsection {Isotropic case}\label{s.isotropic-case}

If the system is statistically isotropic, $\bm{\Phi}(\Delta, t)$ is a
fourth-order isotropic tensor. Moreover, from its definition
[Eq.~(\ref{e.phi-4th-tensor.def})], $\bm{\Phi}(\Delta, t)$ has the following
symmetry properties: $\Phi_{ijkl} = \Phi_{jikl}$, $\Phi_{ijkl} = \Phi_{ijlk}$,
and $\Phi_{ijkl} = \Phi_{klij}$. It follows that $\bm{\Phi}(\Delta, t)$ can be
expressed as
\begin{equation}
  \label{e.phi-4th-tensor-isotropic}
  \Phi_{ijkl}(\Delta, t) =
  \lambda(\Delta, t) \delta_{ij}\delta_{kl} + 
  \mu(\Delta, t) (\delta_{il} \delta_{jk} + \delta_{ik} \delta_{jl}),
\end{equation}
where $\lambda(\Delta, t)$ and $\mu(\Delta, t)$ are scalar functions (these
functions are analogous to the Lam\'e coefficients in the theory of elasticity
for isotropic bodies \cite{landau_elasticity}). Thus, in the isotropic case, the
fourth-order tensor $\bm{\Phi}(\Delta, t)$ is characterized completely by
$\lambda(\Delta, t)$ and $\mu(\Delta, t)$. Taking contractions in
Eq.~(\ref{e.phi-4th-tensor-isotropic}) between the first and second indices
(i.e., $i$ and $j$) and between the third and fourth indices (i.e., $k$ and
$l$), we have
\begin{equation}
  \Phi_1(\Delta, t) = n^2 \lambda(\Delta, t) + 2n \mu(\Delta, t).  
\end{equation}
Similarly, taking contractions between the first and fourth indices (i.e.,
$i$ and $l$) and between the second and third indices (i.e., $j$ and $k$),
we have
\begin{equation}
  \Phi_2(\Delta, t) = n \lambda(\Delta, t) + (n^2+n) \mu(\Delta, t).
\end{equation}
Thus, we reach a significant conclusion that the two scalar functions
$\Phi_1(\Delta, t)$ and $\Phi_2(\Delta, t)$ determine $\bm{\Phi}(\Delta, t)$
entirely for an isotropic system.
For anisotropic systems, however, $\Phi_1(\Delta, t)$ and $\Phi_2(\Delta, t)$
may represent a small part of the information contained in $\bm{\Phi}(\Delta,
t)$. For example, if the spatial dimension $n$ is $3$, as many as 21 elements of
$\bm{\Phi}(\Delta, t)$ are independent.

\subsection {Crossover}

As seen from Eqs.~(\ref{e.phi1}) and (\ref{e.phi2}), the correlation function
$\phi_i(\tau)\,\, (i=1,2)$ satisfies $\lim_{\tau \to \infty} \phi_i(\tau) =
0$. If $\phi_i(\tau)$ has a characteristic time scale $\tau_i$, then, from
Eqs.~(\ref{e.hatPhi1_ex}) and (\ref{e.hatPhi2_ex}), we have
\begin{equation}
  \label{e.Phi_i_crossover}
  \hat{\Phi}_i^{\mathrm{ex}} (t) \approx
  \begin{cases}
    \phi_i (0)                            & (t \ll \tau_i), \\[0.2cm]
    \displaystyle\frac {2}{t}
    \int_{0}^{\infty}\phi_i (\tau)  d\tau & (t \gg \tau_i).
  \end{cases}
\end{equation}
Thus, at the characteristic time scale $\tau_i$, $\hat{\Phi}_i^{\mathrm{ex}}
(t)$ shows a crossover. For the polymer motion studied here, this crossover time
$\tau_i$ corresponds roughly to the longest relaxation time of each polymer
model as shown in the subsequent sections.

\section {Rouse model}\label{s.rouse_model}
%subsection {intro}

In this and the following three sections, the method of the TMSD tensor
developed in the previous section is applied to the four polymer models stated
in the Introduction. Here, the Rouse model is studied as the first example;
although this is a very simple model of a flexible polymer chain in dilute
solutions, it is the basis of many mathematical models of biopolymers
\cite{weber10, gong14, shinkai16}.

%subsection {theory1}

The Rouse model is composed of $N$ equivalent beads, the dynamics of which are
subject neither to the excluded-volume nor hydrodynamic interaction
\cite{rouse53, doi86}:
\begin{equation}
  \label{e.rouse.eqation-of-motion-beads}
  \zeta\frac {\partial \bm{R}_n(t)}{\partial t}
  =
  k\frac {\partial^2 \bm{R}_n(t)}{\partial n^2} + \bm{f}_n(t),
\end{equation}
where $\bm{R}_n(t)$ is the position of bead $n$, $k$ is the spring constant, and
$\zeta$ is the friction coefficient. The spring constant $k$ is related to the
mean bond length $b$ as $k = 3 k_BT / b^2$. The random force $\bm{f}_n(t)$
satisfies $\left\langle \bm{f}_n(t) \right\rangle = 0$ and the
fluctuation-dissipation relation $\left\langle \bm{f}_n(t) \bm{f}_m(t')
\right\rangle = 2\zeta k_BT \delta_{nm}\delta(t-t')\bm{I}$.

The equation of motion for the COM $\bm{R}_{G}(t) := \sum_{n=1}^{N}\bm{R}_n(t)
/N$ is given by
\begin{equation}
  \label{e.rouse.eqation-of-motion-COM}
  \frac {\partial \bm{R}_{G}(t)}{\partial t}
  =
  \sqrt{2 D_G} \bm{\xi}(t),
\end{equation}
where $D_{G} = k_{B}T/N\zeta$ is the diffusion coefficient of the COM. Comparing
with Eq.~(\ref{e.def-lefd}), we have $\bm{B}(t) = \sqrt{D_G}\bm{I}$. Because the
diffusion coefficient $D_G$ is independent of time $t$, we have $\phi_1(t)
\equiv \phi_2(t) \equiv 0$ from Eqs.~(\ref{e.phi1}) and
(\ref{e.phi2}). Consequently, the excess parts also vanish, namely
$\hat{\Phi}_1^{\mathrm{ex}}(t) \equiv \hat{\Phi}_2^{\mathrm{ex}}(t) \equiv 0$,
and, from Eqs.~(\ref{e.hatPhi1_id}) and (\ref{e.hatPhi2_id}), the ideal parts
are given by
\begin{align}
  \label{e.hatPhi1.rouse}
  \hat{\Phi}_1(\Delta, t)
  =%\approx
  \hat{\Phi}_1^{\mathrm{id}}(\Delta, t)
  &=
  \frac {\Delta}{9t} \left(4 - \frac {\Delta}{t}\right),
  \\[0.1cm]
  \label{e.hatPhi2.rouse}
  \hat{\Phi}_2(\Delta, t)
  =%\approx
  \hat{\Phi}_2^{\mathrm{id}}(\Delta, t)
  &=
  \frac {2\Delta}{3t} \left(4 - \frac {\Delta}{t}\right).
\end{align}
Note that the ideal parts decay simply as $1/t$ and do not exhibit crossover.
%% On the other hand, the excess parts exhibit a plateau as shown in the
%% following subsections.

%subsection {fig}

\begin{figure}[t]
  \centerline{\includegraphics[width=8.1cm]{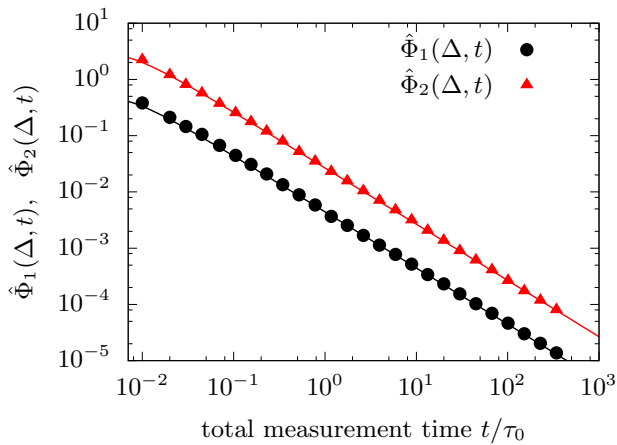}}
  \caption{\label{f.rouse} (color online) Correlation functions
    $\hat{\Phi}_i(t)\,(i=1,2)$ of the TMSD tensor calculated from trajectory
    data $\bm{R}_G(t)$ of the Rouse model (circles and triangles). The COM
    trajectories $\bm{R}_G(t)$ are generated through numerical simulations of
    the Rouse model. Time is measured in units of $\tau_0:= b^2/D$, where $b$ is
    the bond length, and $D$ is the diffusion constant of the beads. The number
    $N$ of beads and the lag time $\Delta$ are set as $N=50$ and $\Delta = 0.01
    \tau_0$. The solid lines are the theoretical predictions given by
    Eqs.~(\ref{e.hatPhi1.rouse}) and (\ref{e.hatPhi2.rouse}). There are no
    fitting parameters (the same is true of
    Figs.~\ref{f.zimm}--\ref{f.rodlike-polymer}).}
\end{figure}

%subsection {simulation}

In Fig.~\ref{f.rouse}, these formulas [Eqs.~(\ref{e.hatPhi1.rouse}) and
(\ref{e.hatPhi2.rouse})] are displayed by the solid lines, and results of the
numerical simulations by the circles and the triangles; the theoretical curves
are in excellent agreement with the simulation results. These numerical results
were obtained from trajectory data $\bm{R}_G(t)$ that were generated through
Brownian dynamics simulations of the Rouse model
[Eq.~(\ref{e.rouse.eqation-of-motion-beads})].

\section {Zimm model}\label{s.zimm_model}
%subsection {intro}

In this section, we study the Zimm model without the excluded volume interaction
(i.e., the Zimm model in the $\Theta$ condition). Some scaling properties of the
Rouse model are known to be inconsistent with experiments \cite{doi86}, which is
because the hydrodynamic interaction
%(which is shown to be important also in macromolecular diffusion in cytoplasms \cite{ando10})
is disregarded entirely in the Rouse model. In contrast, the hydrodynamic
interaction is taken into account in the Zimm model, which is another model of a
flexible polymer chain in dilute solutions.

\subsection {Model definition}

As in the case of the Rouse model, the Zimm model consists of $N$ equivalent
beads, and the equation of motion for bead $n$ is given by \cite{zimm56, doi86,
  ermak78}
%% \begin{equation}
%%   \label{e.def.zimm}
%%   \frac {\partial \bm{R}_n(t)}{\partial t}
%%   =
%%   \sum_m^{} \bm{H}_{nm}  \cdot
%%   \left[
%%   k\frac {\partial^2 \bm{R}_m}{\partial m^2} + \tilde{\bm{f}}_m(t)
%%   \right]
%% \end{equation}
\begin{equation}
  \label{e.def.zimm}
  \frac {\partial \bm{R}_n(t)}{\partial t}
  =
  k\sum_m^{} \bm{H}_{nm} \cdot \frac {\partial^2 \bm{R}_m(t)}{\partial m^2}
  +
  \bm{f}_n(t),
\end{equation}
where the hydrodynamic interaction is represented in terms of the mobility
matrix $\bm{H}_{nm}$ defined by
\begin{align}
  \label{e.H_nn_oseen}
  \bm{H}_{nn} &:= \frac {\bm{I}}{6\pi \eta a},
  \\[0.1cm]
  \label{e.H_nm_oseen}
  \bm{H}_{nm}&:=
  \frac {1}{8\pi\eta r_{nm}}
  \left(\bm{I} + \frac {\bm{r}_{nm}{\bm{r}}_{nm}}{r_{nm}^2}\right)
  \quad (n\neq m).
\end{align}
Here, $\eta$ is the viscosity of the solvent and $a$ is the radius of each
bead. Moreover, $\bm{r}_{nm}(t)$ and $r_{nm}(t)$ are defined as $\bm{r}_{nm}(t)
:= \bm{R}_n(t) - \bm{R}_m(t)$ and $r_{nm}(t) := | \bm{r}_{nm}(t)|$,
respectively. The thermal noise $\bm{f}_m(t)$ satisfies the
fluctuation-dissipation relation
\begin{equation}
  \label{e.zimm.fluctuation-dissipation}
  \left\langle \bm{f}_{n}(t) \bm{f}_{m}(t') \right\rangle
  =
  2k_BT \bm{H}_{nm} \delta(t-t').
\end{equation}
The non-diagonal elements $\bm{H}_{nm}$ [Eq.~(\ref{e.H_nm_oseen})] are known
collectively as the Oseen tensor, the nonlinearity of which makes theoretical
analysis of the Zimm model considerably difficult.

A simple approximation that is commonly adopted is a pre-averaging approximation
\cite{doi86} in which $\bm{H}_{nm}$ is replaced with its equilibrium average
$\left\langle \bm{H}_{nm} \right\rangle =: h(n-m) \bm{I}$. In this
approximation, the equation of motion for bead $n$ is expressed as
\begin{align}
  \label{e.zimm.eq-of-motion.preaveraging}
  \frac {\partial \bm{R}_n(t)}{\partial t}
  \approx
  k\sum_m^{} h(n-m) \frac {\partial^2 \bm{R}_m}{\partial m^2}
  +
  \tilde{\bm{f}}_m(t),
  \\[0.1cm]
  \label{e.zimm.fluctuation-dissipation.preaveraging}
  \bigl\langle \tilde{\bm{f}}_{n}(t) \tilde{\bm{f}}_{m}(t') \bigr\rangle
  =
  2k_BT h (n-m) \bm{I}\delta(t-t') .
\end{align}
Although this approximation works well for predicting the MSD of the COM motion
\cite{rey89}, it is impossible to use it to elucidate the fluctuating
diffusivity. This is because the fluctuating diffusivity is disregarded entirely
when replacing $\bm{H}_{nm}(t)$ in Eq.~(\ref{e.zimm.fluctuation-dissipation})
with $h(n-m)\bm{I}$ [see
Eq.~(\ref{e.zimm.fluctuation-dissipation.preaveraging})].

\subsection {Equation of COM motion}

To elucidate the effect of the fluctuating diffusivity, the pre-averaging
approximation is applied to the internal modes only, whereas the COM motion is
treated without pre-averaging.

The normal mode $\bm{X}_p(t)$ ($p=0,1,\dots$) of $\bm{R}_n(t)$ is defined by \cite{doi86}
\begin{align}
  \label{e.app.zimm.X_p=f(R_n)}
  \bm{X}_p(t) :=&
  \frac {1}{N} \int_{0}^{N} dn
  \cos \left( \frac {p\pi n}{N} \right) \bm{R}_n (t),
  \\[0.1cm]
  \label{e.app.zimm.R_n=f(X_p)}
  \bm{R}_n(t) =&
  \bm{X}_0(t) + 2 \sum_{p=1}^{\infty} 
  \cos \left( \frac {p\pi n}{N} \right)\bm{X}_p(t).
\end{align}
Note here that $\bm{X}_0(t)$ is equivalent to the COM position: $\bm{X}_0(t) =
\bm{R}_G(t)$.  Under the pre-averaging approximation
[Eqs.~(\ref{e.zimm.eq-of-motion.preaveraging}) and
(\ref{e.zimm.fluctuation-dissipation.preaveraging})], the equations of motion
for the normal modes are given by
\begin{align}
  \label{e.zimm.dX_0/dt}
  \frac {\partial \bm{X}_0(t)}{\partial t}
  &=
  \frac {\partial \bm{R}_G(t)}{\partial t}
  =
  \hat{\bm{f}}_0(t),
  \\[0.1cm]
  \label{e.app.zimm.dX/dt.final}
  \frac {\partial \bm{X}_p(t)}{\partial t}
  &=
  -\frac {\bm{X}_p(t)}{\tau_p}
  + \hat{\bm{f}}_p(t)\quad  (p=1,2,\dots),
\end{align}
where $\hat{\bm{f}}_p(t)$ are random forces defined by
\begin{equation}
  \label{e.app.zimm.<f_p(t)f_q(t')>}
  \left\langle
  \hat{\bm{f}}_p(t) \hat{\bm{f}}_q(t')
  \right\rangle
  =
  2k_BT \hat{h}_{pq} \delta(t-t') \bm{I}
  \quad
  (p, q = 0, 1, \dots),
\end{equation}
with
\begin{equation}
  \label{e.zimm.h_pq}
  \hat{h}_{pq}:= \frac {1}{N^2}
  \int_0^N \!\!\!dn
  \int_0^N \!\!\!dm
  \cos \left( \frac {p\pi n}{N} \right)
  \cos \left( \frac {q\pi m}{N} \right)
  h(n-m).
\end{equation}
For $p\neq 0$, $\hat{h}_{pq}$ can be approximated further as $\hat{h}_{pq}
\approx {\delta_{pq}}/{\zeta_p}$, where $\zeta_p:= (12\pi^3 pN)^{1/2} \eta b$
\cite{doi86}. Consequently, the Langevin equations for the internal modes,
Eq.~(\ref{e.app.zimm.dX/dt.final}), are mutually independent because
\begin{equation}
  \label{e.app.zimm.<f_p(t)f_q(t')>.approx}
  \left\langle
  \hat{\bm{f}}_p(t) \hat{\bm{f}}_q(t')
  \right\rangle
  =
  \frac {2k_BT }{\zeta_p} \delta_{pq}\delta(t-t') \bm{I}
  \quad
  (p = 1, 2, \dots).
\end{equation}
Moreover, in Eq.~(\ref{e.app.zimm.dX/dt.final}), $\tau_p$ is the relaxation time
of the $p$-th mode, and given by
\begin{equation}
  \label{e.zimm.tau_p-tau_1}
  \tau_p
  =
  \frac {\eta b^3}{k_BT} \sqrt{\frac {N^3}{3\pi p^3}}
  =
  \frac {\tau_1}{p^{3/2}}\quad  (p=1,2,\dots),
\end{equation}
where $\tau_1$ is the longest relaxation time.

Here, the COM equation of motion in Eq.~(\ref{e.zimm.dX_0/dt}) is rewritten as
\begin{equation}
  \label{e.zimm.dRG(t)/dt}
  \frac {\partial \bm{R}_{G}(t)}{\partial t}
  =
  \sqrt{2 } \bm{B}(t) \cdot \bm{\xi}(t),
\end{equation}
where $\bm{\xi}(t)$ is the white Gaussian noise given by
Eq.~(\ref{e.def-lefd.<xi(t)xi(0)>}). By comparing Eq.~(\ref{e.zimm.dRG(t)/dt})
with Eqs.~(\ref{e.zimm.dX_0/dt}), (\ref{e.app.zimm.<f_p(t)f_q(t')>}), and
(\ref{e.zimm.h_pq}), $\bm{B}(t)$ is given by
\begin{equation}
  \label{e.zimm.D(t)}
  \bm{D}(t) = \bm{B}(t)\cdot\bm{B}^T(t)
  =
  \frac {k_BT}{N^2} \int_{0}^{N}dn\int_{0}^{N}dm \bm{H}_{nm}(t),
\end{equation}
where we restored the time dependence of the diffusivity by formally replacing
$h(n-m) \bm{I}$ with $\bm{H}_{nm}(t)$. In the following analysis,
Eqs.~(\ref{e.app.zimm.dX/dt.final}) and
(\ref{e.app.zimm.<f_p(t)f_q(t')>.approx}) are used for the internal modes,
whereas Eqs.~(\ref{e.zimm.dRG(t)/dt}) and (\ref{e.zimm.D(t)}) are used for
the COM motion.
Thus, the diffusion coefficient of the Zimm model, in contrast to that of the
Rouse model, depends on time $t$ and fluctuates because of the hydrodynamic
interaction.

From Eqs.~(\ref{e.H_nm_oseen}) and (\ref{e.zimm.D(t)}), we have the ensemble
average of the diffusion coefficient tensor as
\begin{align}
  \label{e.zimm.<D>}
  \left\langle \bm{D} \right\rangle
  =&
  \frac {c\bm{I}}{3} 
  %\frac {k_BT \bm{I}}{6\pi \eta N^2}
  \int_{0}^{N}dn\int_{0}^{N}dm
  \left\langle \frac {1}{r_{nm}} \right\rangle,
\end{align}
where $c:= {k_BT}/(2\pi \eta N^2)$ is a constant and we used the mutual
independence of the magnitude $r_{nm}$ and direction $\bm{r}_{nm}/r_{nm}$ as
follows \cite{doi86}:
\begin{align}
  \left\langle \frac {1}{r_{nm}}
  \left(\bm{I} + \frac {\bm{r}_{nm}{\bm{r}}_{nm}} {r_{nm}^2}\right)
  \right\rangle
  %% &=
  %% \left\langle \frac {1}{r_{nm}} \right\rangle
  %% \left\langle
  %% \bm{I} + \frac {\bm{r}_{nm}{\bm{r}}_{nm}} {r_{nm}^2}
  %% \right\rangle
  %% \notag\\[0.1cm]
  &=
  \frac {4}{3}\bm{I}
  \left\langle \frac {1}{r_{nm}} \right\rangle.
\end{align}
From Eq.~(\ref{e.zimm.<D>}), we have
\begin{align}
  \label{e.zimm.tr<D>}
  \mathrm{tr}\left\langle \bm{D} \right\rangle
  &=
  c
  %\frac {k_BT}{2\pi \eta N^2}
  \int_{0}^{N}dn\int_{0}^{N}dm
  \left\langle \frac {1}{r_{nm}} \right\rangle,
  \\[0.1cm]
  \label{e.zimm.tr(<D>.<D>)}
  \mathrm{tr}
  \left(
  \left\langle \bm{D} \right\rangle \cdot
  \left\langle \bm{D} \right\rangle
  \right)
  &=
  \frac {1}{3}
  \left(
  \mathrm{tr}\left\langle \bm{D} \right\rangle
  \right)^2.
  %% \frac {C^2}{3}
  %% % \left(\frac {k_BT}{2\pi \eta N^2}\right)^2
  %% \int_{0}^{N} dn \int_{0}^{N}dm
  %% \left\langle \frac {1}{r_{nm}} \right\rangle
  %% \notag\\[0.1cm]
  %% &\times
  %% \int_{0}^{N} dn'\int_{0}^{N}dm'
  %% \left\langle \frac {1}{r_{n'm'}} \right\rangle,
\end{align}
The validity of Eq.~(\ref{e.zimm.tr<D>}) has been studied intensively
\cite{liu03}, and it is shown that Eq.~(\ref{e.zimm.tr<D>}) is equivalent to the
short-time diffusion coefficient of the COM and that it is also a good
approximation to the long-time diffusion coefficient. In the next subsection,
however, we have to study the second moment of the diffusion coefficient
$\bm{D}(t)$.

Here, $\bm{r}_{nm}(t)$ follows three-dimensional Gaussian distribution with
a covariant matrix $\bm{\Sigma}_3 = \bm{I} |n-m| b^2/3$,
\begin{equation}
  \label{e.gauss-dist-3d}
  f_3(\bm{r})
  =
  \frac {1}{(2\pi |n-m| b^2/3)^{3/2}}
  \exp\left[
  -\frac {1}{2} \bm{r}\cdot \bm{\Sigma}_3^{-1}\cdot \bm{r}
  \right].
\end{equation}
Thus, $\left\langle 1/r_{nm}(t) \right\rangle$ is obtained by integrating over
$f_3(\bm{r})$ in spherical coordinates as \cite{doi86}
\begin{equation}
  \label{e.<1/r>}
  \left\langle \frac {1}{r_{nm}} \right\rangle
  =
  \left(\frac {6}{\pi b^2|n-m|}\right)^{1/2}.
\end{equation}
From Eqs.~(\ref{e.zimm.<D>}) and (\ref{e.<1/r>}), we have an explicit
expression of the ensemble-averaged diffusivity,
\begin{equation}
  \label{e.zimm.<D>.final}
  \left\langle \bm{D} \right\rangle
  =
  \frac {8c}{3b}
  \left(\frac {2N^3}{3\pi}\right)^{1/2}
  \bm{I}.
  %% \frac {4k_BT}{9\pi \eta b}
  %% \left(\frac {6}{\pi N}\right)^{1/2}
  %% \bm{I}.
\end{equation}
It follows that
\begin{align}
  \label{e.zimm.tr<D>.explicit}
  \mathrm{tr} \left\langle \bm{D} \right\rangle
  &=
  \frac {8c}{b}
  \left(\frac {2N^3}{3\pi}\right)^{1/2},
  %% \frac {4k_BT}{3\pi \eta b}
  %% \left(\frac {6}{\pi N}\right)^{1/2},
  \\[0.1cm]
  \label{e.zimm.tr(<D><D>).explicit}
  \mathrm{tr}
  (
  \left\langle \bm{D} \right\rangle \cdot
  \left\langle \bm{D} \right\rangle
  )
  &=
  \left(\frac {8c}{b}\right)^2
  \frac {2N^3}{9\pi}.
  %% \frac {2}{\pi^3 N}
  %% \left(\frac {4k_{B}T}{3\eta b}\right)^2.
\end{align}

\subsection {Correlation functions of diffusion coefficient}
%subsubsection {intro}

In this subsection, we calculate the magnitude and orientation correlation
functions $\phi_1(t)$ and $\phi_2(t)$, respectively, of the diffusivity
[Eqs.~(\ref{e.phi1}) and (\ref{e.phi2})]. In the following derivation, we use
crude approximations such as a single-mode approximation and a perturbation
expansion of the Gaussian distribution. Nevertheless, the final results exhibit
relatively good agreement with those of numerical simulations.

\subsubsection {Magnitude correlation function of diffusion coefficient}

We begin by deriving the magnitude correlation function $\phi_1(t)$ of the
diffusivity. From Eqs.~(\ref{e.H_nm_oseen}) and (\ref{e.zimm.D(t)}), we have
\begin{align}
  \label{e.zimm.<trD(t)trD(0)>}
  &\left\langle
  \mathrm{tr} \bm{D}(t)
  \mathrm{tr} \bm{D}(0)
  \right\rangle
  =
  \notag\\[0.1cm]
  &\hspace*{0.3cm}
  c^2
  \int_{0}^{N}\!\!\!dn \!  \int_{0}^{N}\!\!\!dm\!
  \int_{0}^{N}\!\!\!dn'\!  \int_{0}^{N}\!\!\!dm'\!
  \left\langle \frac {1}{r_{nm}(t)r_{n'm'}(0)} \right\rangle.
\end{align}
To evaluate the ensemble average in the integrand, we define a six-dimensional
vector $\bm{X} := (x, x', y,y', z, z')$,
%% \begin{equation}
%%   \label{e.zimm.X}
%%   \bm{X} := (x, x', y, y', z, z'),  
%% \end{equation}
where $(x, y, z) := \bm{r}_{nm}(t)$ and $(x', y', z') := \bm{r}_{n'm'}(0)$.  It
can be shown that $\bm{X}$ follows six-dimensional Gaussian distribution (see
Appendix \ref{s.app.covariant-matrix} for a derivation), namely
\begin{equation}
  \label{e.gauss-dist-6d}
  f_6(\bm{X})
  =
  \frac {1}{(2\pi)^3 \left|\bm{\Sigma}_6 \right|^{1/2}}
  \exp\left[
  -\frac {1}{2} \bm{X}\cdot \bm{\Sigma}_6^{-1}\cdot \bm{X}
  \right].
\end{equation}
Here, $\bm{\Sigma}_6$ is a $6\times6$ covariant matrix defined by
\begin{equation}
  \bm{\Sigma}_6
  :=
  \begin{pmatrix}
    \bm{A} & \bm{0} & \bm{0} \\[0.cm]
    \bm{0} & \bm{A} & \bm{0} \\[0.cm]
    \bm{0} & \bm{0} & \bm{A} \\[0.cm]
  \end{pmatrix},
  %% \begin{pmatrix}
  %%   \alpha   & \beta    & \BigZero & \BigZero  \\[-.2cm]
  %%   \beta    & \alpha'  &                      \\[-.0cm]
  %%   \BigZero & \alpha   & \beta    & \BigZero  \\[-.2cm]
  %%   &          & \beta    & \alpha' & \\[-.0cm]
  %%   \BigZero & \BigZero & \alpha   & \beta     \\[-.2cm]
  %%   &          &          &         & \beta & \alpha' 
  %% \end{pmatrix}.
  \qquad
  \bm{A}
  :=
  \begin{pmatrix}
    \alpha & \beta \\
    \beta  & \alpha'
  \end{pmatrix},
\end{equation}
where $\bm{0}$ is the $2\times 2$ zero matrix; $\alpha, \alpha'$ and $\beta$ are
defined by (see Appendix \ref{s.app.covariant-matrix})
\begin{align}
  \label{e.zimm.alpha}
  \alpha  =& \frac {b^2}{3}|n-m|, \qquad   
  \alpha' = \frac {b^2}{3}|n'-m'|, \\[0.1cm]
  \beta
  =&
  \frac {8Nb^2}{3\pi^2}
  \sum_{p=1}^{\infty}
  \frac {e^{-t/\tau_p}}{p^2}
  \sin \frac {p\pi(n+m)}{2N}
  \sin \frac {p\pi(n-m)}{2N}
  \notag\\[-0.1cm]
  \label{e.zimm.beta}
  &\hspace*{1.7cm}\times
  \sin \frac {p\pi(n'+m')}{2N}
  \sin \frac {p\pi(n'-m')}{2N}.
\end{align}
Hereinafter, we take only the longest relaxation mode ($p=1$) into account and
ignore all the other modes (i.e., a single-mode approximation):
\begin{align}
  \label{e.beta.approx}
  \beta
  \approx
  \frac {8Nb^2}{3\pi^2}
  e^{-t/\tau_1}&
  \sin \frac {\pi(n+m)}{2N}
  \sin \frac {\pi(n-m)}{2N}
  \notag\\[0.1cm]
  \times&
  \sin \frac {\pi(n'+m')}{2N}
  \sin \frac {\pi(n'-m')}{2N}.
\end{align}
Consequently, the determinant of the covariant matrix $\bm{\Sigma}_6$ is given
by
\begin{equation}
  \left| \bm{\Sigma}_6 \right|
  =
  \left(\alpha\alpha' - \beta^2\right)^3
  =
  \left(\alpha\alpha'\right)^3
  \left(1 - \epsilon\right)^3
  =
  \left(\tilde{\alpha}\alpha'\right)^3,
\end{equation}
where $\epsilon := \beta^2/(\alpha\alpha')$ and $\tilde{\alpha} :=
\alpha(1-\epsilon)$.

Using these quantities in Eq.~(\ref{e.gauss-dist-6d}),
we have
\begin{align}
  \left\langle \frac {1}{r_{nm}(t)r_{n'm'}(0)} \right\rangle
  &=
  \notag\\[0.1cm]
  \label{e.<1/r(t)r'(0)>}
  &\hspace*{-2.3cm}
  \frac {\tilde{\alpha}\tilde{\alpha}'}{(2\pi)^3 \left| \bm{\Sigma}_6\right|^{1/2}}
  \int d\bm{r} \int d\bm{r}' 
  \frac {1}{rr'}
  e^{
    -\frac {r^2}{2}
    -\frac {r'^2}{2}
    + \epsilon^{1/2} \bm{r}\cdot\bm{r}'},
\end{align}
where $\bm{r}:= (x,y,z)$, and $\bm{r}':= (x',y',z')$.  For $t \gg \tau_1$, we
have $\epsilon \ll 1$ and the above integrand can be approximated further as
\begin{align}
  \label{e.6d-gauss-approx}
  e^{\epsilon^{1/2} \bm{r}\cdot\bm{r}'}\simeq
  \quad 1 +
  \epsilon^{1/2} \bm{r}\cdot\bm{r}' +
  \frac {\epsilon}{2} (\bm{r}\cdot\bm{r}')^2.
\end{align}
Integrating Eq.~(\ref{e.<1/r(t)r'(0)>}) in spherical coordinates, we have a
perturbation expansion upto order $\epsilon^1$ as
\begin{align}
  \left\langle \frac {1}{r_{nm}(t)r_{n'm'}(0)} \right\rangle
  &\simeq
  \frac {2}{\pi (\alpha\alpha')^{1/2}}
  \left(1+ \frac {\epsilon}{6}\right)
  \notag\\[0.1cm]
  &=
  \left\langle \frac {1}{r_{nm}} \right\rangle
  \left\langle \frac {1}{r_{n'm'}} \right\rangle
  +
  \frac {\epsilon}{3\pi (\alpha\alpha')^{1/2}},
\end{align}
where we used Eqs.~(\ref{e.<1/r>}) and (\ref{e.zimm.alpha}). Inserting this
equation into Eq.~(\ref{e.zimm.<trD(t)trD(0)>}) and taking
Eq.~(\ref{e.zimm.tr<D>}) into account, we obtain
\begin{align}
  \label{e.zimm.<trD(t)trD(0)>.final}
  \left\langle
  \mathrm{tr} \bm{D}(t)
  \mathrm{tr} \bm{D}(0)
  \right\rangle
  &
  -
  \left(
  \mathrm{tr}
  \left\langle \bm{D} \right\rangle
  \right)^2
  \notag\\[0.1cm]
  =&
  \frac {c^2}{3\pi}
  \int_{0}^{N}\!\!\!dn\!
  \int_{0}^{N}\!\!\!dm\!
  \int_{0}^{N}\!\!\!dn'\!
  \int_{0}^{N}\!\!\!dm'
  \frac {\epsilon}{(\alpha\alpha')^{1/2}}
  \notag\\[0.1cm]
  =&
  \left(\frac {8c}{b}\right)^2
  \frac {N^3K^2}{\pi^6}
  e^{-2t/\tau_1},
\end{align}
where $K$ is a constant defined by
\begin{equation}
  K :=
  \int_{0}^{\pi} d \xi\,
  \frac {\sin^2 \frac {\xi}{2}}{\xi^{3/2}}
  \left(
  \pi - \xi + \sin\xi
  \right)
  \approx 1.428226.
\end{equation}
Finally, from Eqs.~(\ref{e.zimm.tr<D>.explicit}) and
(\ref{e.zimm.<trD(t)trD(0)>.final}), we have the magnitude correlation function
$\phi_1(t)$ of the diffusivity [Eq.~(\ref{e.phi1})] as
\begin{equation}
  \label{e.zimm.phi1}
  \phi_1(t) = \frac {3K^2}{2\pi^5} e^{-2t/\tau_1}.
\end{equation}

\subsubsection {Orientation correlation function of diffusion coefficient}

We move on to a derivation of the orientation correlation function $\phi_2(t)$
of the diffusivity [Eq.~(\ref{e.phi2})]. From Eqs.~(\ref{e.H_nm_oseen}) and
(\ref{e.zimm.D(t)}), we have
\begin{align}
  \mathrm{tr}
  \left\langle
  \bm{D}(t)\cdot \bm{D}(0)
  \right\rangle
  = &
  \frac {c^2}{16}
  %% \left(
  %% \frac {k_BT}{8\pi\eta N^2}
  %% \right)^2
  \int_{0}^{N}\!\!\!dn\!
  \int_{0}^{N}\!\!\!dm\!
  \int_{0}^{N}\!\!\!dn'\!
  \int_{0}^{N}\!\!\!dm'
  \notag\\[0.1cm]
  \label{e.zimm.tr<D(t).D(0)>}
  &
  \times
  \left\langle
  \frac
  {\left[\hat{\bm{r}}_{nm}(t)\cdot \hat{\bm{r}}_{n'm'}(0)\right]^2 + 5}
  {r_{nm}(t)r_{n'm'}(0)}
  \right\rangle,
\end{align}
where $\hat{\bm{r}}$ is the unit vector in the direction of $\bm{r}$.  The
ensemble average in Eq.~(\ref{e.zimm.tr<D(t).D(0)>}) can be carried out in a way
similar to the calculation of Eq.~(\ref{e.<1/r(t)r'(0)>}). In fact, under the
approximation in Eq.~(\ref{e.6d-gauss-approx}), we obtain
\begin{align}
  \left\langle
  \frac
  {\left[\hat{\bm{r}}_{nm}(t)\cdot \hat{\bm{r}}_{n'm'}(0)\right]^2 + 5}
  {r_{nm}(t)r_{n'm'}(0)}
  \right\rangle
  &=
  \frac {16}{3}
  \frac {2}{\pi(\alpha\alpha')^{1/2}}
  \left(1+\frac {1}{5}\epsilon\right)
  \notag\\[0.1cm]
  &\hspace*{-2.5cm}=
  \label{e.zimm.tr<D(t).D(0)>.integrand}
  \frac {16}{3} 
  \left\langle \frac {1}{r_{nm}} \right\rangle
  \left\langle \frac {1}{r_{n'm'}} \right\rangle
  +
  \frac {32}{15} 
  \frac {\epsilon}{\pi (\alpha\alpha')^{1/2}}.
\end{align}
Inserting Eq.~(\ref{e.zimm.tr<D(t).D(0)>.integrand}) into
Eq.~(\ref{e.zimm.tr<D(t).D(0)>}) and taking Eqs.~(\ref{e.zimm.tr<D>}) and
(\ref{e.zimm.tr(<D>.<D>)}) into account, we have
\begin{align}
  \mathrm{tr}
  \left\langle
  \bm{D}(t)\cdot \bm{D}(0)
  \right\rangle
  &-
  \mathrm{tr}
  \left(
  \left\langle \bm{D} \right\rangle \cdot
  \left\langle \bm{D} \right\rangle
  \right)
  \notag\\[0.1cm]
  =&
  \frac {2c^2}{15\pi} 
  \int_{0}^{N}\!\!\!dn\!
  \int_{0}^{N}\!\!\!dm\!
  \int_{0}^{N}\!\!\!dn'\!
  \int_{0}^{N}\!\!\!dm'
  \frac {\epsilon}{(\alpha\alpha')^{1/2}}
  \notag\\[0.1cm]
  \label{e.zimm.tr<D(t).D(0)>.final}
  =&
  \frac {2}{5}
  \left(\frac {8c}{b}\right)^2
  \frac {N^3K^2}{\pi^6}
  e^{-2t/\tau_1},
\end{align}
Finally, from Eqs.~(\ref{e.zimm.tr(<D><D>).explicit}) and
(\ref{e.zimm.tr<D(t).D(0)>.final}), we have the orientation correlation function
$\phi_2(t)$ of the diffusivity [Eq.~(\ref{e.phi2})] as
\begin{align}
  \label{e.zimm.phi2}
  \phi_2(t) &= \frac {9K^2}{5\pi^5} e^{- 2t/\tau_1}.
\end{align}

%subsection {figure: Phi1 & Phi2}
\begin{figure}[]
  \centerline{\includegraphics[width=8.1cm]{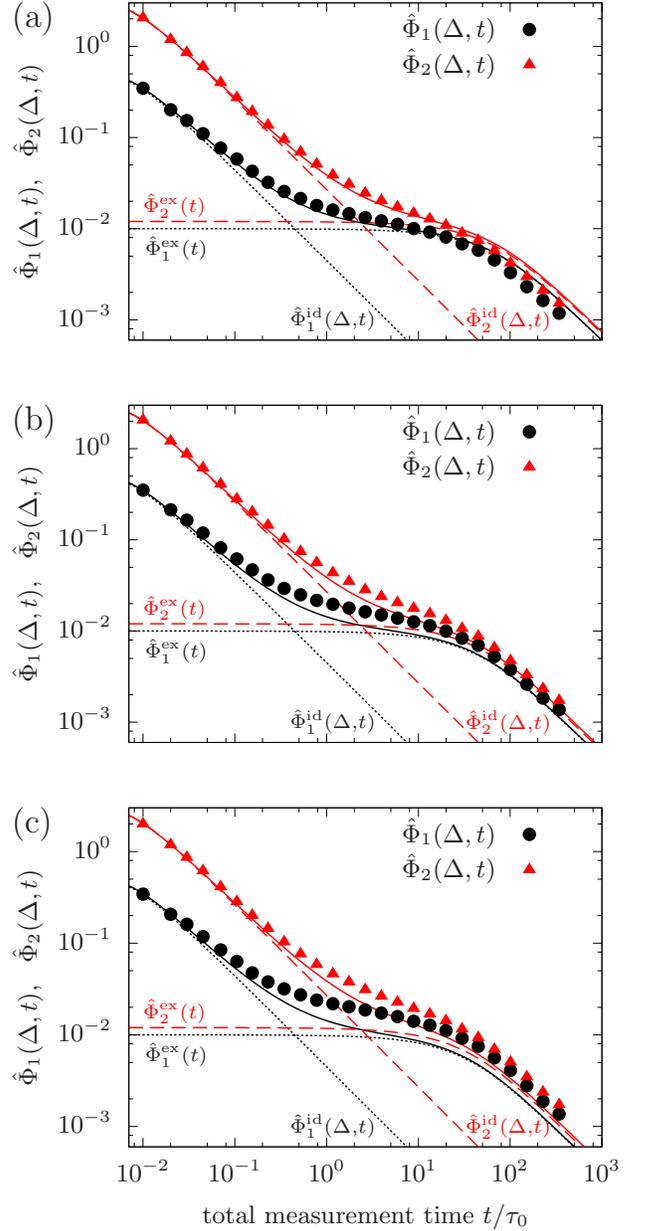}}
  %% \resizebox{81mm}{!}{\input{rsd_zimm_RDS010}}\vspace*{-.0cm}\\
  %% \resizebox{81mm}{!}{\input{rsd_zimm_RDS015}}\vspace*{-.0cm}\\
  %% \resizebox{81mm}{!}{\input{rsd_zimm_RDS020}}
  \caption{\label{f.zimm} (color online) Correlation functions
    $\hat{\Phi}_i(t)\,(i=1,2)$ of the TMSD tensor calculated from trajectory
    data $\bm{R}_G(t)$ of the Zimm model (circles and triangles). The COM
    trajectories $\bm{R}_G(t)$ are generated through numerical simulations of
    the Zimm model (see Appendix \ref{s.rotne-prager-yamakawa-tensor}).
    Distance is measured in units of the bond length $b$ and time in units of
    $\tau_0 := b^2/D$, where $D$ is the diffusion coefficient of each
    bead. Results for three different values of the bead radius $a$ are
    presented: (a) $a = 0.1b$, (b) $a = 0.15b$ and (c) $a = 0.2b$. The number
    $N$ of beads and the lag time $\Delta$ are set as $N=50$ and $\Delta =
    0.01\tau_0$. The longest relaxation time $\tau_1$ is estimated from
    Eq.~(\ref{e.zimm.tau_p-tau_1}) as (a) $\tau_1 = 61.0\tau_0$, (b) $\tau_1 =
    40.7\tau_0$, and (c) $\tau_1 = 30.5\tau_0$. The dotted lines are the
    theoretical predictions for $\hat{\Phi}_1^{\mathrm{id}}(\Delta, t)$ and
    $\hat{\Phi}_1^{\mathrm{ex}}(t)$ given by Eqs.~(\ref{e.zimm_Phi1^id.final})
    and (\ref{e.zimm_Phi1^ex.final}). The dashed lines are the theoretical
    predictions for $\hat{\Phi}_2^{\mathrm{id}}(\Delta, t)$ and
    $\hat{\Phi}_2^{\mathrm{ex}}(t)$ given by Eqs.~(\ref{e.zimm_Phi2^id.final})
    and (\ref{e.zimm_Phi2^ex.final}). The solid lines are the sums
    $\hat{\Phi}_i(\Delta, t)\,\, (i=1,2)$ of the ideal and excess parts
    [Eqs.~(\ref{e.hatPhi_1=hatPhi_1^id+hatPhi_1^ex}) and
    (\ref{e.hatPhi_2=hatPhi_2^id+hatPhi_2^ex})].}
\end{figure}

\subsection {Correlation functions of TMSD tensor}

Here, we derive the correlation functions $\hat{\Phi}_{1}(\Delta, t)$ and
$\hat{\Phi}_{2}(\Delta, t)$ of the TMSD tensor.  From Eqs.~(\ref{e.hatPhi1_id}),
(\ref{e.hatPhi1_ex}), (\ref{e.zimm.phi1}), and (\ref{e.zimm.phi2}), we have
\begin{align}
  \label{e.zimm_Phi1^id.final}
  \hat{\Phi}_{1}^{\mathrm{id}}(\Delta, t)
  &=
  \frac {\Delta}{9t} \left(4 - \frac {\Delta}{t}\right)
  \left(1 + \frac {9K^2}{5\pi^5}\right),
  \\[0.1cm]
  \label{e.zimm_Phi1^ex.final}
  \hat{\Phi}_{1}^{\mathrm{ex}}(\Delta, t)
  &=
  \frac {3K^2\tau_1^2}{4\pi^5t^2}
  \left[
  \frac {2t}{\tau_1} + e^{-2t/\tau_1} - 1 
  \right],
\end{align}
where we used $C=1$ because the system is statistically isotropic.  Similarly,
from Eqs.~(\ref{e.hatPhi2_id}), (\ref{e.hatPhi2_ex}), (\ref{e.zimm.phi1}), and
(\ref{e.zimm.phi2}), we obtain
\begin{align}
  \label{e.zimm_Phi2^id.final}
  \hat{\Phi}_2^{\mathrm{id}}(\Delta, t)
  &=
  \frac {\Delta}{6t} \left(4 - \frac {\Delta}{t}\right)
  \left(4 + \frac {63K^2}{10\pi^5}\right),  
  \\[0.1cm]
  \label{e.zimm_Phi2^ex.final}
  \hat{\Phi}_{2}^{\mathrm{ex}}(\Delta, t)
  &=
  \frac {6}{5} \hat{\Phi}_{1}^{\mathrm{ex}}(\Delta, t). 
  %% \frac {9K^2\tau_1}{5\pi^5t}
  %% \left[
  %% 1 + \frac {\tau_1}{2t}e^{-2t/\tau_1} - \frac {\tau_1}{2}
  %% \right]
\end{align}

In contrast to the Rouse model, these correlation functions
$\hat{\Phi}_i^{\mathrm{ex}}(t)$ $(i=1,2)$ for the Zimm model show
crossovers. For example, from Eq.~(\ref{e.Phi_i_crossover}),
$\hat{\Phi}_1^{\mathrm{ex}}(t)$ behaves as
\begin{align}
  \label{e.phi1.zimm.asympt}
  \hat{\Phi}_1^{\mathrm{ex}} (t) \simeq
  \begin{cases}
    \frac {3K^2}{2\pi^5}  & (t \ll \tau_1), \\[0.1cm]
    \frac {3K^2 \tau_1}{2\pi^5t} & (t \gg \tau_1).
  \end{cases}
\end{align}
From Eq.~(\ref{e.phi1.zimm.asympt}), the crossover time $\tau_c$ can be
estimated as $\tau_c = \tau_1$, i.e., the crossover time is equivalent to the
longest relaxation time $\tau_1$. Also, $\hat{\Phi}_2^{\mathrm{ex}}(t)$ shows a
crossover at $t=\tau_1$, because of Eq.~(\ref{e.zimm_Phi2^ex.final}).

As can be seen in Fig.~\ref{f.zimm}, the theoretical predictions (the solid
lines) [Eqs.~(\ref{e.zimm_Phi1^id.final})--(\ref{e.zimm_Phi2^ex.final})] are in
good agreement with the results of the numerical simulations (the symbols). The
slight deviations are due to the approximations used in the theoretical
analysis. For example, in the simulations, the Rotne-Prager-Yamakawa tensor
[Eq.~(\ref{e.rotne-prager-yamakawa-tensor})] was utilized as the mobility matrix
instead of the Oseen tensor [Eq.~(\ref{e.H_nm_oseen})] to regularize the
singularity in the Oseen tensor at $r_{nm} = 0$. Moreover, we also applied the
pre-averaging approximation to the inner degrees of freedom, and used the
perturbation expansion in Eq.~(\ref{e.6d-gauss-approx}). However, incorporating
a higher order term ($\epsilon^2$) in Eq.~(\ref{e.6d-gauss-approx}) improves the
theoretical predictions only slightly (its contribution is less than 15 \% of
the leading term; data not shown).

%% simulation results for three different values of the bead radius, $a$, are
%% displayed (circles and triangles), and they exhibit qualitatively similar to
%% the theoretical curves (solid lines). For example, the simulation curves have
%% plateaus in intermediate time regimes and exhibit crossovers at around
%% $t=\tau_1$ as predicted by the theory.

%% the simulation results and the theoretical curves show better agreement for
%% smaller values of $a$. This may be due to the fact that theoretical analysis
%% in the preceding subsections is carried out for the Oseen tensor, which
%% formally corresponds to the Rotne-Prager-Yamakawa tensor with $a \to 0$.

\section {Discrete reptation model}\label{s.reptation_model}
%subsection {intro}

In this section, the focus is on the discrete reptation model, which describes
tagged polymer motion in entangled polymer solutions \cite{doi78,
  doi86}. Because of the entanglement, the tagged polymer chain of the reptation
model is temporarily trapped in a virtual tube comprised of surrounding chains,
and moves only in the longitudinal direction of the tube. Such reptation
dynamics are an essential ingredient in modeling DNA molecules at high
concentration \cite{gong14}.

%subsection {theory1}

In the reptation model, the centerline of the tube, which is called a primitive
chain, is considered instead of the real chain of the tagged polymer. The
primitive chain is assumed to consist of $N$ tube segments $\bm{R}_1(t),\dots,
\bm{R}_N(t)$ connected by bonds of constant length $b$. The primitive chain is
allowed to move only in the longitudinal direction of the tube as a result of
the entanglement. A single step of the primitive-chain dynamics is given as
follows; one of the two end segments, $\bm{R}_1(t)$ or $\bm{R}_N(t)$, is chosen
with equal probability; the chosen end segment hops with step length $b$ in a
random direction; and each of the other segments slides to one of the positions
of its neighboring segments [i.e., if $\bm{R}_1(t)$ is chosen, $\bm{R}_{n}(t)$
slides to $\bm{R}_{n-1}(t)$ ($n=2,\dots,N$); if $\bm{R}_N(t)$ is chosen,
$\bm{R}_{n}(t)$ slides to $\bm{R}_{n+1}(t)$ ($n=1,\dots,N-1$)].

The COM $\bm{R}_{G}(t)$ of this primitive chain follows the LEFD
[Eq.~(\ref{e.def-lefd})] with $\bm{B}(t)$ given by \cite{doi78, uneyama15}
\begin{equation}
  \bm{B}(t)
  \approx
  \sqrt{\frac {3D_{G}}{\left\langle \bm{p}^2 \right\rangle}}
  \frac {\bm{p}(t) \bm{p}(t)}{|\bm{p}(t)|},
\end{equation}
where $D_G$ is the ensemble-averaged diffusion coefficient of the COM, and
$\bm{p}(t)$ is the end-to-end vector of the primitive chain. It follows that
the diffusion coefficient is obtained from Eq.~(\ref{e.D(t)=B(t).B^T(t)}) as
\begin{equation}
  \label{e.reptation.D(t)}
  \bm{D}(t)
  =
  3 D_G
  \frac {\bm{p}(t) \bm{p}(t)}{\left\langle \bm{p}^2 \right\rangle}.
\end{equation}
Because the system is statistically isotropic, $\left\langle \bm{p}\bm{p}
\right\rangle = A \bm{I}$ with a constant $A$. Taking the trace, we have $A =
\left\langle \bm{p}^2 \right\rangle/3$. It follows that $\left\langle \bm{D}
\right\rangle = D_G \bm{I}$.

By using Eqs.~(\ref{e.phi1}), (\ref{e.phi2}) and (\ref{e.reptation.D(t)}), the
magnitude and the orientation correlation functions $\phi_1(\tau)$ and
$\phi_2(\tau)$ of the diffusivity can be expressed as
\begin{align}
  \label{e.phi1.reptation}
  \phi_{1}(\tau)
  &=
  \frac
  {\left\langle \bm{p}^2(\tau) \bm{p}^2(0) \right\rangle}
  {\left\langle \bm{p}^2 \right\rangle^2}
  - 1,
  \\[0.1cm]
  \label{e.phi2.reptation}
  \phi_{2}(\tau)
  &=
  3\frac
  {\langle \left[\bm{p}(\tau)\cdot \bm{p}(0)\right]^2 \rangle}
  {\left\langle \bm{p}^2 \right\rangle^2}
  - 1.
\end{align}
In Ref.~\cite{uneyama15}, $\phi_1(\tau)$ was obtained explicitly as
\begin{equation}
  \phi_{1}(\tau)
  =
  \frac {16}{3\pi^2}
  \sum_{k: \mathrm{odd}}^{}
  \frac 1{k^2}
  E_2\left(\frac {k^2 \tau}{\tau_d}\right),
\end{equation}
where $\tau_d$ is the longest relaxation time of the reptation model, and
$E_m(x)$ is the generalized exponential integral of order $m$ \cite{olver10}.
Furthermore, it is shown in Appendix \ref{s.app.reptation} that
\begin{equation}
  \label{e.reptation.phi_2=6phi_1}
  \phi_{2}(\tau) = 6 \phi_{1} (\tau).
\end{equation}

%subsection {fig}

\begin{figure}[t]
  \centerline{\includegraphics[width=8.1cm]{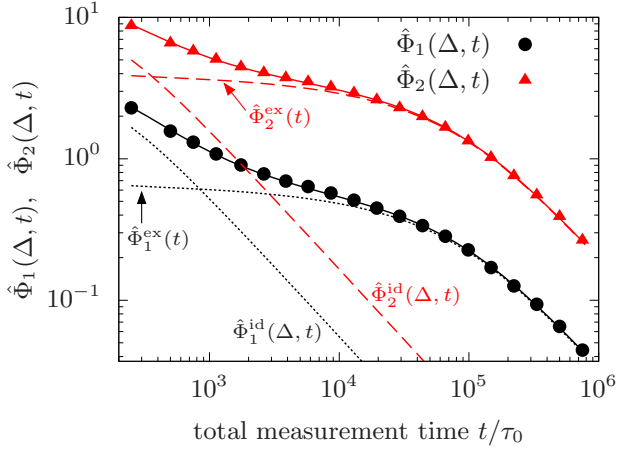}}
  %\resizebox{81mm}{!}{\input{rsd_reptation}}
  \caption{\label{f.reptation} (color online) Correlation functions
    $\hat{\Phi}_i(t)\,(i=1,2)$ of the TMSD tensor calculated from trajectory
    data $\bm{R}_G(t)$ of the reptation model (circles and triangles). The COM
    trajectories $\bm{R}_G(t)$ are generated through numerical
    simulations of the reptation model. Distance is measured in units of the
    bond length $b$ between the tube segments, and time in units of $\tau_0 :=
    b^2\zeta / k_BT$, where $\zeta$ is a friction coefficient of the tube
    segment.  %
    The number $N$ of tube segments and the lag time $\Delta$ are set as
    $N=85$ and $\Delta = 250 \tau_0$. For details of the simulation setup, see
    Ref.~\cite{uneyama12}. The longest relaxation time (a disengagement time)
    $\tau_d = N^3\tau_0/\pi^2$ is estimated as $\tau_d \approx 6.2\times 10^4
    \tau_0$.
    The dotted lines are the theoretical predictions for
    $\hat{\Phi}_1^{\mathrm{id}}(\Delta, t)$ and $\hat{\Phi}_1^{\mathrm{ex}}(t)$
    given by Eqs.~(\ref{e.reptation.Phi1^id}) and
    (\ref{e.reptation.Phi1^ex}). The dashed lines are the theoretical
    predictions for $\hat{\Phi}_2^{\mathrm{id}}(\Delta, t)$ and
    $\hat{\Phi}_2^{\mathrm{ex}}(t)$ given by Eqs.~(\ref{e.reptation.Phi2^id})
    and (\ref{e.reptation.Phi2^ex}). The solid lines are the sums
    $\hat{\Phi}_i(\Delta, t)\,\, (i=1,2)$ of the ideal and excess parts
    [Eqs.~(\ref{e.hatPhi_1=hatPhi_1^id+hatPhi_1^ex}) and
    (\ref{e.hatPhi_2=hatPhi_2^id+hatPhi_2^ex})].}
\end{figure}

%subsection {theory2}

From Eqs.~(\ref{e.hatPhi1_id}) and (\ref{e.hatPhi1_ex}), we have the correlation
functions of the TMSD tensor as
\begin{align}
  \label{e.reptation.Phi1^id}
  \hat{\Phi}_{1}^{\mathrm{id}}(\Delta, t)
  &=
  \frac {5\Delta}{9t} \left(4 - \frac {\Delta}{t}\right),
  \\[0.1cm]
  \label{e.reptation.Phi1^ex}
  \hat{\Phi}_{1}^{\mathrm{ex}}(\Delta, t)
  &=
  \frac {\pi^2\tau_d}{18t} -
  \frac {\pi^4 \tau_d^2}{270t^2} +
  \frac {32\tau_d^2}{3\pi^2 t^2}\sum_{k: \mathrm{odd}}^{}
  \frac {E_4\left(\frac {k^2t}{\tau_d}\right)}{k^6},
\end{align}
where we used $\phi_2(0) = 6\phi_1(0) = 4$.  Similarly, from
Eqs.~(\ref{e.hatPhi2_id}), (\ref{e.hatPhi2_ex}) and
(\ref{e.reptation.phi_2=6phi_1}), we have
\begin{align}
  \label{e.reptation.Phi2^id}
  \hat{\Phi}_{2}^{\mathrm{id}}(\Delta, t)
  &=
  \frac {5\Delta}{3t} \left(4 - \frac {\Delta}{t}\right)
  =
  3 \hat{\Phi}_{1}^{\mathrm{id}}(\Delta, t), 
  \\[0.1cm]
  \label{e.reptation.Phi2^ex}
  \hat{\Phi}_{2}^{\mathrm{ex}}(\Delta, t)
  &=
  6\hat{\Phi}_{1}^{\mathrm{ex}}(\Delta, t).
\end{align}

%subsection {theory3}

As in the case of the Zimm model, both functions $\hat{\Phi}_i^{\mathrm{ex}}(t)$
$(i=1,2)$ show crossovers. For example, $\hat{\Phi}_1^{\mathrm{ex}}(t)$ behaves
as \cite{uneyama15}
\begin{align}
  \label{e.phi1.reptation.asympt}
  \hat{\Phi}_1^{\mathrm{ex}} (t) \simeq
  \begin{cases}
    \frac {2}{3}              & (t \ll \tau_{d}), \\[0.1cm]
    \frac {\pi^2 \tau_d}{18t} & (t \gg \tau_{d}).
  \end{cases}
\end{align}
Also, $\hat{\Phi}_2^{\mathrm{ex}}(t)$ shows a crossover at $t=\tau_d$, because
of Eq.~(\ref{e.reptation.Phi2^ex}). From Eq.~(\ref{e.phi1.reptation.asympt}), 
this crossover time can be estimated as
\begin{equation}
  \label{e.reptation.t_c}
  \tau_c = \frac {\pi^2}{12} \tau_d,
\end{equation}
which is close to the longest relaxation time $\tau_d$.

In Fig.~\ref{f.reptation}, results of the numerical simulations for the discrete
reptation model are displayed; they exhibit remarkable agreement with the
theoretical predictions
[Eqs.~(\ref{e.reptation.Phi1^id})--(\ref{e.reptation.Phi2^ex})].
%% Both
%% $\hat{\Phi}_1(\Delta, t)$ and $\hat{\Phi}_2(\Delta, t)$ show crossover around at
%% $\tau_c$.
Moreover, $\hat{\Phi}_2^{\mathrm{ex}}(\Delta, t)$ far exceeds
$\hat{\Phi}_1^{\mathrm{ex}}(\Delta, t)$ in the reptation model, in contrast to
the Zimm model for which the two functions are comparable. This means that the
orientation fluctuation of the diffusivity is more prominent in the reptation
model than in the Zimm model.

\section {Rigid rod-like polymer}\label{s.rodlike_polymer}
%subsection {intro}

Finally, the rigid rod-like polymer in a dilute condition is investigated as an
extreme example of non-flexible polymers \cite{%burgers1938error-for-PRE,
  burgers95, doi86, dhont96, berne00}. In general, it is more difficult to
observe rotational diffusion of an anisotropic particle than it is to observe
its translational diffusion \cite{han06}. With the TMSD tensor analysis,
however, the rotational diffusion coefficient can be estimated by measuring
translational motion of the COM.

%subsection {theory1}

Let us denote the COM of the rod as $\bm{R}_{G}(t)$, and assume that the rod is
cylindrically symmetric along the long axis. Consequently, the COM position
$\bm{R}_{G}(t)$ follows the LEFD [Eq.~(\ref{e.def-lefd})] with $\bm{B}(t)$ given
(see Appendix \ref{s.app.rod}) by
\begin{equation}
  \label{e.rodlike-polymer.B(t)}
  \bm{B}(t) =
  \sqrt{D_{\parallel}} \hat{\bm{u}}(t)\hat{\bm{u}}(t) +
  \sqrt{D_{\perp}} \left[ \bm{I} - \hat{\bm{u}}(t)\hat{\bm{u}}(t)\right],
\end{equation}
where $\hat{\bm{u}}(t)$ is a unit vector in the direction of the rod's long
axis, and $D_{\parallel}$ and $D_{\perp}$ are the diffusion coefficients along
and perpendicular to the long axis, respectively. Moreover, it is assumed that
the rod is long and thin so that rotational motion around the long axis is
disregarded.
The time evolution of the rod's direction $\hat{\bm{u}}(t)$ is given by
\cite{dhont96}
\begin{align}
  \label{e.rodlike-polymer.du/dt}
  \frac {d\hat{\bm{u}}(t)}{dt}
  &=
  \sqrt{2D_r}
  \bm{\eta}(t) \times \hat{\bm{u}}(t),
  \\[0.1cm]
  \label{e.rodlike-polymer.<eta(t)eta(t')>}
  \left\langle \bm{\eta}(t) \right\rangle
  =
  0,
  &\quad
  \left\langle \bm{\eta}(t)\bm{\eta}(t') \right\rangle
  =
  \bm{I} \delta(t-t'),
\end{align}
where $\bm{\eta} (t)$ is white Gaussian noise, and $D_r$ is the rotational
diffusion coefficient. The three diffusion coefficients $D_{\parallel}$,
$D_{\perp}$ and $D_{r}$ can be expressed in terms of the length $L$ and diameter
$b$ of the rod as \cite{doi86, dhont96}
\begin{align}
  \label{e.rod.D_parallel.D_perp}
  D_{\parallel} = 2 D_{\perp} &= \frac {k_BT \ln(L/b)}{2 \pi \eta L},
  \\[0.1cm]
  \label{e.rod.D_r}
  D_{r} &= \frac {3 k_BT \ln(L/b)}{ \pi\eta L^3}. 
\end{align}
These formulas are obtained through hydrodynamic calculations for a long thin
rod, i.e., $L/b \gg 1$.

Firstly, we consider the magnitude correlation function $\hat{\Phi}_1(t)$. From
Eqs.~(\ref{e.D(t)=B(t).B^T(t)}) and (\ref{e.rodlike-polymer.B(t)}), we have the
fluctuating diffusivity as
\begin{equation}
  \label{e.rod.D(t)}
  \bm{D}(t)
  =
  D_{\parallel} \hat{\bm{u}}(t)\hat{\bm{u}}(t) +
  D_{\perp} \left[ \bm{I} - \hat{\bm{u}}(t)\hat{\bm{u}}(t)\right].
\end{equation}
Taking the trace, we obtain $\mathrm{tr} \bm{D}(t) = D_{\parallel} + 2
D_{\perp}$, i.e., the magnitude of the diffusivity is constant in time. It
follows that the magnitude correlation of the diffusivity vanishes, i.e.,
$\phi_1(\tau) \equiv 0$; hence we have from Eq.~(\ref{e.hatPhi1_ex}) that
\begin{equation}
  \label{e.rod.Phi1^ex}
  \hat{\Phi}_1^{\mathrm{ex}}(t) \equiv 0.
\end{equation}
Thus, for the rigid rod-like polymer, in contrast to the Zimm and reptation
models, it is impossible to extract information about the fluctuating
diffusivity by using $\hat{\Phi}_1(t)$ .

%subsection {fig}

\begin{figure}[t]
  \centerline{\includegraphics[width=8.1cm]{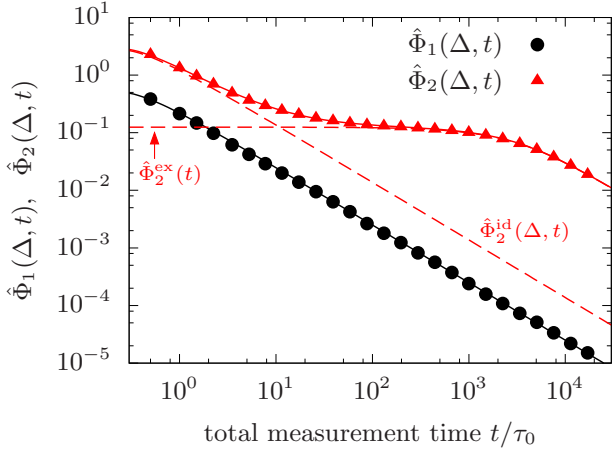}}
  %\resizebox{81mm}{!}{\input{rsd_rod}}
  \caption{\label{f.rodlike-polymer} (color online) Correlation functions
    $\hat{\Phi}_i(t)\,(i=1,2)$ of the TMSD tensor calculated from trajectory
    data $\bm{R}_G(t)$ of the rigid rod-like polymer (circles and
    triangles). The COM trajectories $\bm{R}_G(t)$ are generated through
    numerical simulations of the rigid rod-like polymer
    [Eqs.~(\ref{e.def-lefd}), (\ref{e.rodlike-polymer.B(t)}), and
    (\ref{e.rodlike-polymer.du/dt})]. Distance is measured in units of the rod's
    diameter $b$ and time in units of $\tau_0 := b^3\eta/k_BT$.  The length $L$
    of the rod is set as $L = 30b$, and $\Delta$ is $0.5\tau_0$.  The values of
    the three diffusion coefficients ($D_{\perp}, D_{\parallel}, D_r$) are given
    by Eqs.~(\ref{e.rod.D_parallel.D_perp}) and (\ref{e.rod.D_r}).  In
    particular, the rotational relaxation time $1/D_r$ is estimated from
    Eq.~(\ref{e.rod.D_r}) as $1/D_r \approx 8.3 \times 10^3 \tau_0$.
    The dashed lines are the theoretical predictions for
    $\hat{\Phi}_2^{\mathrm{id}}(\Delta, t)$ and $\hat{\Phi}_2^{\mathrm{ex}}(t)$
    given by Eqs.~(\ref{e.rod.Phi2^id}) and (\ref{e.rod.phi2^ex}). The solid
    lines are the sums $\hat{\Phi}_i(\Delta, t)\,\, (i=1,2)$ of the ideal and
    excess parts [Eqs.~(\ref{e.hatPhi_1=hatPhi_1^id+hatPhi_1^ex}) and
    (\ref{e.hatPhi_2=hatPhi_2^id+hatPhi_2^ex})].}
\end{figure}

%subsection {theory2}

Therefore, to elucidate the fluctuating diffusivity of the rod, it is necessary
to study $\hat{\Phi}_2(t)$.  From Eqs.~(\ref{e.phi2}) and (\ref{e.rod.D(t)}), we
obtain the orientation correlation function as
\begin{align}
  \phi_2(\tau)
  =
  \left(\frac {D_{\parallel} -D_{\perp}}{D_{\parallel} + 2 D_{\perp}}\right)^2
  \left\{
  3\left\langle
  \bigl[\hat{\bm{u}}(\tau)\cdot\hat{\bm{u}}(0)\bigr]^2
  \right\rangle
  - 1
  \right\},
\end{align}
where we used $\left\langle \bm{D} \right\rangle = (D_{\parallel} + 2 D_{\perp})
\bm{I} / 3$.  Because the rotational motion given by
Eq.~(\ref{e.rodlike-polymer.du/dt}) is independent of the translational motion
$\bm{R}_G(t)$, the correlation function $\langle
[\hat{\bm{u}}(\tau)\cdot\hat{\bm{u}}(0)]^2 \rangle$ can be calculated by
employing the Smoluchowsky equation for the rotational motion as \cite{doi86,
  berne00}
\begin{equation}
  \left\langle
  \bigl[\hat{\bm{u}}(\tau)\cdot\hat{\bm{u}}(0)\bigr]^2
  \right\rangle
  =
  \frac {1}{3} \left(1 + 2 e^{-6D_r t}\right),
\end{equation}
and hence we have
\begin{align}
  \phi_2(\tau)
  =
  2
  \left(\frac {D_{\parallel} -D_{\perp}}{D_{\parallel} + 2 D_{\perp}}\right)^2
  e^{-6 D_r \tau}.
\end{align}
From Eq.~(\ref{e.hatPhi2_ex}), the excess part $\hat{\Phi}_2^{\mathrm{ex}}(t)$
of the orientation correlation function is obtained as
\begin{align}
  \label{e.rod.phi2^ex}
  \hat{\Phi}_2^{\mathrm{ex}}(t)
  =
  \frac {1}{9 D_r^2t^2}
  \left(\frac {D_{\parallel} -D_{\perp}}{D_{\parallel} + 2 D_{\perp}}\right)^2
  \left(
  6D_r t + e^{-6D_r t} - 1
  \right).
\end{align}
Moreover, by using Eqs.~(\ref{e.hatPhi1_id}) and (\ref{e.hatPhi2_id}), the ideal
parts are given by
\begin{align}
  \label{e.rod.Phi1^id}
  \hat{\Phi}_1^{\mathrm{id}}(t)
  &=
  \frac {\Delta}{9t} \left(4 - \frac {\Delta}{t}\right)
  \left[1 +
  2\left(\frac {D_{\parallel} -D_{\perp}}{D_{\parallel} + 2 D_{\perp}}\right)^2
  \right],
  \\[0.1cm]
  \label{e.rod.Phi2^id}
  \hat{\Phi}_2^{\mathrm{id}}(t)
  &=
  \frac {\Delta}{3t} \left(4 - \frac {\Delta}{t}\right)
  \left[
  2 +
  \left(\frac {D_{\parallel} -D_{\perp}}{D_{\parallel} + 2 D_{\perp}}\right)^2
  \right],
\end{align}
where $C=1$ is used because the system is statistically isotropic.
In particular, from Eq~(\ref{e.Phi_i_crossover}), we have the following
crossover:
\begin{align}
  \label{e.phi2.rodlike-polymer.asympt}
  \hat{\Phi}_2^{\mathrm{ex}}(t) \approx
  \begin{cases}
    2 \left(\frac {D_{\parallel} -D_{\perp}}{D_{\parallel} + 2 D_{\perp}}\right)^2 & (t \ll 1/D_r),
    \\[0.1cm]
    \frac {2}{3 D_{r}t}
    \left(\frac {D_{\parallel} -D_{\perp}}{D_{\parallel} + 2 D_{\perp}}\right)^2   & (t \gg 1/D_r).
  \end{cases}
\end{align}
An estimate for the rotational relaxation time $1/D_r$ can be obtained
from this crossover time, despite the fact that we observe only the
translational motion of the rod. In fact, we have the crossover time $\tau_c$
from Eq.~(\ref{e.phi2.rodlike-polymer.asympt}) as
\begin{equation}
  \tau_c = \frac {1}{3 D_r}.
\end{equation}
Thus, the crossover time $\tau_c$ gives an estimate of the rotational relaxation
time $1/D_r$.

%subsection {simulation}

Results of the numerical simulations for $\hat{\Phi}_{1}(t)$ and
$\hat{\Phi}_{2}(t)$ are presented in Fig.~\ref{f.rodlike-polymer} (the circles
and triangles). As predicted, $\hat{\Phi}_{1}(t)$ shows no crossover because the
excess part is absent, whereas $\hat{\Phi}_{2}(t)$ exhibits a clear
crossover. The numerical results are consistent with the theoretical predictions
(the solid lines).

%% クロスオーバーがレプテーションモデルの場合よりクリアなのは, レプテーションモデル
%% の場合は最長緩和時間に対応する緩和モード以外にも複数の内部自由度を持つのに対し,
%% 剛体棒状ポリマーの場合は内部自由度として回転の緩和モードしか存在しないことによる
%% と考えられる.

\section {Discussion}\label{s.discussion}
%subsubsection {summary of general background}

The sample-to-sample fluctuation of the diffusivity observed both in SPT
experiments and theoretical models has been studied intensively for a decade. In
such studies, the sample-to-sample fluctuation is usually characterized by the
EB parameter \cite{he08, deng09, jeon11, tabei13, metzler14, miyaguchi11c,
  miyaguchi13, thiel14b, thiel14, uneyama12, uneyama15, miyaguchi15a,
  miyaguchi16}. However, when calculating the EB parameter from trajectory data
$\bm{r}(t)$, much of the information originally contained in the data is
lost. In this study, to obtain more information from the trajectory data, the EB
parameter is generalized into the fourth-order tensor $\bm{\Phi}(\Delta, t)$,
which is a correlation function of the TMSD tensor. Moreover, the two scalar
functions $\hat{\Phi}_1(\Delta, t)$ and $\hat{\Phi}_2(\Delta, t)$ are derived
from $\bm{\Phi}(\Delta, t)$; these functions are closely related to the
magnitude and orientation correlation functions of the diffusivity, and in
particular $\hat{\Phi}_1(\Delta, t)$ is equivalent to the EB parameter.  It is
also worth noting that a linear combination of the excess parts
$\hat{\Phi}_1^{\mathrm{ex}}(\Delta, t)$ and $\hat{\Phi}_2^{\mathrm{ex}}(\Delta,
t)$ gives the non-Gaussian parameter $A(t)$
[Eq.~(\ref{e.non-gaussian-parameter})]. In other words, the non-Gaussianity can
be decomposed into two parts: one originating from the magnitude fluctuation of
the diffusivity, and the other from the orientation fluctuation.

%% physical meaning of the EB parameter is not clear, because theoretical
%% studies have been conducted mainly for simple models such as the CTRW and
%% fractional Brownian motions. Moreover, it is desirable if we have more
%% information from trajectory data other than the EB parameter.

%subsubsection {summary of this article, achievement/cotribution}

Furthermore, by using the TMSD tensor analysis, it is shown that the four
polymer models exhibit distinctly different types of fluctuating diffusivity in
terms of the correlation functions $\hat{\Phi}_1^{\mathrm{ex}}(\Delta, t)$ and
$\hat{\Phi}_2^{\mathrm{ex}}(\Delta, t)$. For example,
$\hat{\Phi}_1^{\mathrm{ex}}(\Delta, t) \approx
\hat{\Phi}_2^{\mathrm{ex}}(\Delta, t)$ in the Zimm model,
$\hat{\Phi}_1^{\mathrm{ex}}(\Delta, t) < \hat{\Phi}_2^{\mathrm{ex}}(\Delta, t)$
in the reptation model, and $\hat{\Phi}_1^{\mathrm{ex}}(\Delta, t) \ll
\hat{\Phi}_2^{\mathrm{ex}}(\Delta, t)$ in the rigid rod-like polymer. This is in
contrast to the non-Gaussian parameter $A(t)$, whose behavior is qualitatively
similar for these three models; hence the polymer models are barely
distinguishable with $A(t)$.

From these results, it seems that the fluctuating diffusivity might be
ubiquitous in polymer motions from dilute to concentrated solutions and from
flexible to non-flexible polymers. This is because the Zimm and the reptation
models are flexible polymer models in dilute and concentrated solutions,
respectively; in contrast, the rigid rod-like polymer is an extreme case of
non-flexible polymers; each of these three models exhibits fluctuating
diffusivity.

Moreover, the rotational relaxation time $\tau_r = 1/D_r$ of the rigid rod can
be obtained from the crossover time of $\hat{\Phi}_2^{\mathrm{ex}}(t)$
[Eq.~(\ref{e.phi2.rodlike-polymer.asympt})]. As a more direct approach, $\tau_r
= 1/D_r$ of an anisotropic particle was obtained in Ref.~\cite{han06} by
measuring the particle's direction. Also, with the results of
Refs.~\cite{brenner63, cichocki12}, $\tau_r$ of the rigid rod can be estimated
from the ensemble-averaged MSD of a reference point on the rod other than its
COM. For both methods, however, it is necessary to measure at least one
reference point other than the COM. In contrast, with the method proposed here,
$\tau_r$ can be estimated by measuring only the translational motion of the COM.

Of course, the same information of $\hat{\Phi}_1^{\mathrm{ex}}(t)$ and
$\hat{\Phi}_2^{\mathrm{ex}}(t)$ would be obtained from the ensemble-averaged
quantities. In fact, the functions $\hat{\Phi}_1^{\mathrm{ex}}(t)$ and
$\hat{\Phi}_2^{\mathrm{ex}}(t)$ are related to the non-Gaussian parameter $A(t)$
[Eq.~(\ref{e.non-gaussian-parameter})], which is defied by a fourth
moment. Thus, essentially the same information as
$\hat{\Phi}_1^{\mathrm{ex}}(t)$ and $\hat{\Phi}_2^{\mathrm{ex}}(t)$ might well
be obtained from the translational correlation tensor of fourth order, which
might be analyzed by the traditional approach with the Smoluchowski equation
\cite{rallison78, cichocki12, cichocki15}. However, It should be noted that to
calculate fourth moments such as $A(t)$ accurately, a large number of
trajectories are necessary in general.
%% But, in the single-particle-tracking experiments,
%% only tens or hundreds of trajectories are available in general.
In contrast, the present method, in which the time and ensemble averages are
combined, works for a relatively small number of trajectories (typically, from
tens to hundreds of trajectories), and therefore it would be useful in
single-particle-tracking experiments, in which much effort is required to obtain
a large number of trajectories.
%% It is shown that COM motions of these models can be described by the Langevin
%% equation with fluctuating diffusivity [Eq.~(\ref{e.def-lefd})].

%% In addition, the crossover in these functions gives the longest relaxation time
%% of the inner degrees of freedom of polymers. In particular, it is interesting
%% that, for the rigid rod-like polymer, we can obtain the rotational relaxation
%% time only from the measurements of translational COM motions.

%subsubsection {limitations, future work, applications}

Although the TMSD tensor analysis for the polymer models is based on the fact
that the COM of these models can be described in terms of the LEFD
[Eq.~(\ref{e.def-lefd})], there are many phenomena that cannot be described with
the LEFD. For example, the motion of a single bead in the Zimm and reptation
models does not follow the LEFD because the bead shows anomalous subdiffusion,
whereas the LEFD exhibits only normal diffusion as shown in
Eq.~(\ref{e.et_msd_tensor}). A candidate for describing such complex dynamics
might be a generalized Langevin equation or fractional Brownian motion with
fluctuating diffusivity, but the physical validity of such models should be
clarified in future work.

Moreover, only two scalar functions, namely $\hat{\Phi}_1(\Delta, t)$ and
$\hat{\Phi}_2(\Delta, t)$, were used here to analyze the isotropic polymer
models. However, there must still be useful information in the fourth-order
tensor $\bm{\Phi}(\Delta, t)$ for the case of anisotropic systems (see
Sec.~\ref{s.isotropic-case}). Future work should therefore include a full
characterization of this tensor $\bm{\Phi}(\Delta, t)$.

%% All the polymer models studied in this article is statistically isotropic, but
%% the present method can be applied to anisotropic systems (e.g., Brownian motion
%% in a shear flow \cite{dhont96}).

%% いることがしばしば報告されている \cite{richert02}. このような動的不均一
%% 性 (dynamical heterogeneity) においても, 単一分子の拡散係数は揺らいでいること
%% が予想される \cite{pastore15}. あるいは, ブラウン粒子系の流体力学的相互作用
%% も, 拡散性の揺らぎとみなすことができるかもしれない \cite{doi86, dhont96}. そこ
%% で, このような系に対して TMSD 解析を行うことによって, 系に関する重要な情報が得
%% られることが期待される.

%section {acknowledgments}

\begin{acknowledgments}
  The author would like to thank T.~Akimoto and T.~Uneyama for fruitful
  discussions and comments. This work was supported by JSPS KAKENHI for Young
  Scientists (B) (Grant No. JP15K17590).
\end{acknowledgments}

\appendix {}
\section {Decomposition of fourth-order tensor $\bm{\Phi}(\Delta, t)$ into ideal and excess parts}
%{Derivation of Eq.~(\ref{e.phi-4th-tensor.phi_id+phi_ex})}
\label{s.derivation.Phi=Phi_id+Phi_ex}

In this appendix, the expression for $\bm{\Psi}^1(\Delta, t)$ given in
Eq.~(\ref{e.final.Psi1}) is derived. First, using Eqs.~(\ref{e.def-lefd}) and
(\ref{e.tmsd-tensor}), we obtain
\begin{align}
  \label{e.integral-expression.Psi1}
  \bm{\Psi}^1(\Delta, t)
  \simeq
  \frac {1}{t^2}
  \int_0^t dt'\int_0^t dt''
  \bm{\Omega}(\Delta, t', t''),
\end{align}
where $t-\Delta$ is approximated as $t-\Delta \simeq t$, and
$\bm{\Omega}(\Delta, t', t'')$ is another fourth-order tensor defined by
\begin{widetext}
  \begin{align}
    \label{e.def.Omega}
    \bm{\Omega}(\Delta, t', t'')
    &=
    4
    \int_{t'}^{t'+\Delta}   ds
    \int_{t'}^{t'+\Delta}   ds'
    \int_{t''}^{t''+\Delta} du
    \int_{t''}^{t''+\Delta} du'
    \left\langle
    \bm{B} (s)\cdot \bm{\xi}(s)\,
    \bm{B} (s')\cdot \bm{\xi}(s')\,
    \bm{B} (u)\cdot \bm{\xi}(u)\,
    \bm{B} (u')\cdot \bm{\xi}(u')
    \right\rangle.
  \end{align}
  By using the Heaviside step function $\Theta(t)$ and Wick's theorem, namely
  \begin{align}
    \label{e.wick}
    \left\langle \xi_j(s)\xi_l(s')\xi_n(u)\xi_q(u')
    \right\rangle
    =
    \delta_{jl}\delta_{nq}\delta(s-s')\delta(u-u') +
    \delta_{jn}\delta_{lq}\delta(s-u)\delta(s'-u') +
    \delta_{jq}\delta_{ln}\delta(s-u')\delta(s'-u),
  \end{align}
  the elements of
  $\bm{\Omega}(\Delta, t', t'')$ for $t'' < t'$ is obtained as
  \begin{equation}
    \label{e.final-expression.Omega}
    \Omega_{ikmp}(\Delta, t', t'')
    \approx
    4
    \left(\left\langle
    D_{im}D_{kp} \right\rangle + \left\langle D_{ip}D_{km}
    \right\rangle\right)
    \Theta(t''+\Delta-t') \left(t'' + \Delta - t'\right)^2
    +
    4\Delta^2 \left\langle D_{ik}(t') D_{mp}(t'') \right\rangle,
  \end{equation}
\end{widetext}
where approximations such as $D_{ik}(s) \approx D_{ik}(t')$ for $s\in [t',
t'+\Delta)$ are applied; these approximations are justified by the assumption
that $\Delta$ is much shorter than a characteristic time scale of the
fluctuating diffusivity $\bm{D}(t)$.
In addition, an expression similar to Eq.~(\ref{e.final-expression.Omega}) can
be obtained also for $t' < t''$.
By putting these equations into Eq.~(\ref{e.integral-expression.Psi1}) and using
the stationarity, the elements of $\bm{\Psi}^1(\Delta, t)$ can be expressed as
Eq.~(\ref{e.final.Psi1}).

\section {Derivation of six-dimensional covariant matrix $\bm{\Sigma}_{6}$ for Zimm model}
\label{s.app.covariant-matrix}

Here, the covariant matrix $\bm{\Sigma}_{6}$ of the six-dimensional Gaussian
distribution in Eq.~(\ref{e.gauss-dist-6d}) is derived.
Firstly, let us denote a transition probability density function (PDF) for the
normal mode $\bm{X}_p(t)$ as $P_p(\bm{X}_p,t | \bm{X}'_{p},0)$; more precisely,
$P_p(\bm{X}_p,t | \bm{X}'_{p},0) d\bm{X}_p$ is the transition probability from
$\bm{X}'_p$ at time $0$ to an interval $[\bm{X}_p, \bm{X}_p + d\bm{X}_p)$ at
time $t$.
From Eqs.~(\ref{e.app.zimm.dX/dt.final}) and
(\ref{e.app.zimm.<f_p(t)f_q(t')>.approx}), the PDF for $\bm{X}_p(t) \,
(p=1,2,\dots)$ is given by
\begin{equation}
  \label{e.app.zimm.P(X_p|X'_p)}
  P_p(\bm{X}_p,t | \bm{X}'_{p},0)
  =
  \frac {1}{ [2\pi\sigma_p^2(t)]^{3/2} }
  \exp\left[
  -
  \frac
  {(\bm{X}_{p} - \bm{X}'_p e^{-t/\tau_p})^2}
  {2 \sigma^2_p(t)}
  \right],
\end{equation}
where
%$\sigma_p^2(t) = (k_BT/k_p) (1-e^{-2t/\tau_p})$
$\sigma_p^2(t) = (k_BT \tau_p/\zeta_p) (1-e^{-2t/\tau_p})$ is the variance. For
example, Eq.~(\ref{e.app.zimm.P(X_p|X'_p)}) can be derived by using
Chandrasekhar's theorem \cite{dhont96}. In particular, by taking the limit $t\to
\infty$, the equilibrium PDF for $\bm{X}_p$ is obtained as
\begin{equation}
  P_p^{\mathrm{eq}}(\bm{X}_p)
  =
  \frac {1}{(2\pi\sigma_p^2)^{3/2}}
  \exp\left(
  -
  \frac
  {\bm{X}_{p}^2}
  {2 \sigma^2_p}
  \right),
\end{equation}
where %$\sigma_p^2 = k_BT/k_p$. %
$\sigma_p^2 = k_BT \tau_p/\zeta_p$.

By using these PDFs, the joint PDF of $\bm{r}_{nm}(t)$ and $\bm{r}_{n'm'}(0)$ is
expressed as
\begin{widetext}
  \begin{equation}
    \label{e.app.zimm.P(r,t;r',0)}
    P(\bm{r},t; \bm{r}',0)
    =
    \int 
    \int 
    \delta\left( \sum_{p=1}^{\infty} c_p \bm{X}_{p} -\bm{r}\right)
    \delta\left( \sum_{p=1}^{\infty} c_p' \bm{X}'_{p} -\bm{r}'\right)
    \prod_{p=1}^{\infty}
    P_p(\bm{X}_p,t | \bm{X}'_{p},0)
    P_p^{\mathrm{eq}}(\bm{X}'_p)
    d\bm{X}_p
    d\bm{X}_p',
  \end{equation}
\end{widetext}
where $P(\bm{r},t; \bm{r}',0)d\bm{r}d\bm{r}'$ is the joint probability that
$\bm{r}_{nm}(t)$ is in $[\bm{r}, \bm{r}+d\bm{r})$ and $\bm{r}_{n'm'}(0)$ is in
$[\bm{r}', \bm{r}'+d\bm{r}')$.
Note here that $\bm{r}_{nm}(t)$ and $\bm{r}_{n'm'}(0)$ are written as
$\bm{r}_{nm}(t) = \sum_{p=1}^{\infty} c_p \bm{X}_p(t)$ and
$\bm{r}_{n'm'}(0) = \sum_{p=1}^{\infty} c'_p \bm{X}_p(0)$ with
\begin{align}
c_p &:= -4
\sin \frac {p\pi(n+m)}{2N}
\sin \frac {p\pi(n-m)}{2N},
\\[0.1cm]
c'_p &:= -4
\sin \frac {p\pi(n'+m')}{2N}
\sin \frac {p\pi(n'-m')}{2N},
\end{align}
because of Eq.~(\ref{e.app.zimm.R_n=f(X_p)}) and $ \bm{r}_{nm}(t) =
\bm{R}_n(t)-\bm{R}_m(t)$.

With Fourier transformation of Eq.~(\ref{e.app.zimm.P(r,t;r',0)}) with respect
to $\bm{r}$ and $\bm{r}'$, we have a characteristic function
\begin{equation}
  \label{e.app.zimm.P(k;k')}
  \hat{P}(\bm{k},t; \bm{k}',0)
  =
  \exp\left[
  -\frac {1}{2} \alpha \bm{k}^2
  -\frac {1}{2} \alpha' \bm{k}'^2
  - \beta \bm{k}\cdot \bm{k}'
  \right],
\end{equation}
where $\alpha$, $\alpha'$, and $\beta$ are defined in Eqs. (\ref{e.zimm.alpha})
and (\ref{e.zimm.beta}). Moreover, $\bm{k}$ and $\bm{k}'$ are the Fourier
variables conjugate to $\bm{r}$ and $\bm{r}'$, respectively; their elements are
defined as $\bm{k} := (k_x, k_y, k_z)$ and $\bm{k}' := (k'_x, k'_y, k'_z)$.  To
derive Eq.~(\ref{e.app.zimm.P(k;k')}), we used a Fourier series
$\sum_{p=1}^{\infty} \cos (px) /p^2 = (x-\pi)^2/4 - \pi^2/12$ for $x\in [0,
2\pi]$.
If we define a variable $\bm{K}$ as $\bm{K} := (k_x, k_x', k_y, k_y', k_z,
k_z')$, the right-hand side of Eq.~(\ref{e.app.zimm.P(k;k')}) can be rewritten
as
\begin{equation}
  \hat{P}(\bm{k},t; \bm{k}',0)
  =
  \exp\left[
  -\frac {1}{2} \bm{K} \cdot \bm{\Sigma}_6 \cdot \bm{K}
  \right]
  =:
  \hat{P}(\bm{K}).
\end{equation}
This is a characteristic function of six-dimensional Gaussian distribution;
consequently, Fourier inversion of $\hat{P}(\bm{K})$ gives
Eq.~(\ref{e.gauss-dist-6d}).

\section {Rotne-Prager-Yamakawa tensor}
\label{s.rotne-prager-yamakawa-tensor}

To carry out numerical simulations of the Zimm model, it is necessary to
regularize the singularity of the Oseen tensor at $r_{nm}=0$
[Eq.~(\ref{e.H_nm_oseen})].  A commonly employed regularization method is the
Rotne--Prager--Yamakawa tensor $\tilde{\bm{H}}_{nm} \,(n \neq m)$ \cite{rotne69,
  yamakawa70}:
\begin{align}
  \label{e.rotne-prager-yamakawa-tensor}
  \tilde{\bm{H}}_{nm} =
  \begin{cases}
    &\frac {1}{8\pi\eta r_{nm}}
    \left[
    \left(\bm{I} + \frac {\bm{r}_{nm}{\bm{r}}_{nm}}{r_{nm}^2}\right)
    +
    \frac {2a^2}{r_{nm}^2}
    \left(\frac {\bm{I}}{3} - \frac {\bm{r}_{nm}{\bm{r}}_{nm}}{r_{nm}^2} \right)
    \right]\\[0.1cm]
    & \hfill (r_{nm} \geq 2a),
    \\[0.2cm]
    &\frac {1}{6\pi\eta a}
    \left[
    \left( 1 - \frac {9}{32}\frac {r_{nm}}{a}\right) \bm{I}
    +
    \frac {3}{32} \frac {\bm{r}_{nm}{\bm{r}}_{nm}}{r_{nm} a}
    \right]\\[0.1cm]
    & \hfill  (r_{nm} < 2a),
  \end{cases}
\end{align}
where $a$ is the bead radius. In our numerical simulations for the Zimm model,
$\tilde{\bm{H}}_{nm}$ was used for the mobility matrix $\bm{H}_{nm}$ in
Eq.~(\ref{e.def.zimm}).  The Langevin equation [Eq.~(\ref{e.def.zimm})] was
solved numerically by using the Ermak--McCammon algorithm \cite{ermak78}.

\section {Derivation of correlation functions for discrete reptation model}
\label{s.app.reptation}

In this Appendix, the relation presented in Eq.~(\ref{e.reptation.phi_2=6phi_1})
is derived for the discrete reptation model. The end-to-end vector $\bm{p}(t)$
of the reptation model can be expressed with a bond vector $\bm{u}(s,t)$ as
\begin{equation}
  \bm{p} (t) = \int_0^N ds \bm{u}(s,t),
\end{equation}
where $s$ is the segment index and $N$ is the number of segments.  The bond
vector $\bm{u}(s,t)$ follows a Gaussian distribution with zero mean, and any two
bond vectors $\bm{u}(s,t)$ and $\bm{u}(s',t)$ are mutually independent. Thus,
the first and second moments of $\bm{u}(s,t)$ in equilibrium are given by
\begin{equation}
  \label{e.app.reptation.<u>_and_<uu>}
  \left\langle \bm{u}(s) \right\rangle =0,\qquad
  \left\langle \bm{u}(s) \bm{u}(s') \right\rangle
  =
  \frac {b^2}{3}\delta(s-s') \bm{I}, 
\end{equation}
where $b$ is the bond length.

To derive an explicit formula for $\phi_1(\tau)$ and $\phi_2(\tau)$
[Eqs.~(\ref{e.phi1.reptation}) and (\ref{e.phi2.reptation})], we use the
survival probability $\Psi (s; t)$ of segment $s$; more precisely, $\Psi (s; t)$
is the probability that segment $s$ at time $0$ survives until time $t$
\cite{doi78}. Also, we define a survival joint probability $\Psi (s,s'; t)$ of
two segments $s$ and $s'$ \cite{uneyama15}. Namely, $\Psi (s,s'; t)$ is the
probability that both segments $s$ and $s'$ at time $0$ survive until time
$t$. In particular, $\Psi (s,s; t) = \Psi(s;t)$ is satisfied.  Although an
explicit expression for $\Psi (s,s'; t)$ was derived in Ref.\cite{uneyama15}, it
is not required here.

Correlation functions of the end-to-end vector $\bm{p}(t)$ can be expressed with
$\bm{u}(s,t)$. For example, a fourth-order correlation function (tensor) of
$\bm{p}(t)$ is written as
\begin{widetext}
  \begin{align}
    \label{e.app.reptation.<p(t)p(t)p(0)p(0)>}
    \left\langle \bm{p}(t)\bm{p}(t)\bm{p}(0)\bm{p}(0) \right\rangle
    &=
    \int_{0}^{N}ds\int_{0}^{N}ds'
    \int_{0}^{N}dv\int_{0}^{N}dv'
    \left\langle
    \bm{u}(s,t) \bm{u}(s',t)
    \bm{u}(v,0) \bm{u}(v',0)
    \right\rangle.
  \end{align}
  The elements of the tensor in the integrand can be rewritten as
  \begin{align}
    \left\langle
    u_i(s,t)u_j(s',t)
    u_k(v,0)u_l(v',0)
    \right\rangle
    =&
    \left\langle
    u_i(s)u_j(s')
    u_k(v)u_l(v')
    \right\rangle
    \Psi (s,s'; t)
    \notag\\[0.1cm]
    &+
    \left\langle u_i(s) \right\rangle
    \left\langle
    u_j(s')
    u_k(v)
    u_l(v')
    \right\rangle
    \left[\Psi (s';t) - \Psi(s,s';t)\right]
    \notag\\[0.1cm]
    &+
    \left\langle u_j(s') \right\rangle
    \left\langle
    u_i(s)
    u_k(v)
    u_l(v')
    \right\rangle
    \left[\Psi (s;t) - \Psi(s,s';t)\right]
    \notag\\[0.1cm]
    \label{e.app.reptation.<u(t)u(t)u(0)u(0)>}
    &+
    \left\langle u_i(s) u_j(s') \right\rangle
    \left\langle u_k(v) u_l(v') \right\rangle
    \left[1 - \Psi (s;t) - \Psi (s';t) + \Psi(s,s';t)\right],
  \end{align}
  where $\Psi (s';t) - \Psi(s,s';t)$ is the probability that only segment $s'$
  survives, and $1 - \Psi (s;t) - \Psi (s';t) + \Psi(s,s';t)$ is the probability
  that neither of segments $s$ and $s'$ survive.  By using
  Eq.~(\ref{e.app.reptation.<u>_and_<uu>}), the second and third terms on the
  right-hand side vanish.  Meanwhile, the ensemble averages in the first and
  fourth terms can be rewritten as
  \begin{align}
    \label{e.app.reptation.<uuuu>}
    \left\langle
    u_i(s)u_j(s')
    u_k(v)u_l(v')
    \right\rangle
    =&
    \frac {b^4}{9}
    \left[
    \delta_{ij}\delta_{kl} \delta(s-s') \delta(v-v') +
    \delta_{ik}\delta_{jl} \delta(s-v) \delta(s'-v') +
    \delta_{il}\delta_{jk}  \delta(s-v') \delta(s'-v)
    \right],
    \\[0.1cm]
    \label{e.app.reptation.<uu><uu>}
    \left\langle u_i(s)  u_j(s') \right\rangle
    \left\langle u_k(v) u_l(v') \right\rangle
    =&
    \frac {b^4}{9}
    \delta_{ij}\delta_{kl} 
    \delta(s-s') \delta(v-v'),
  \end{align}
  where we used Wick's theorem \cite{doi86} and
  Eq.~(\ref{e.app.reptation.<u>_and_<uu>}).  Putting
  Eqs.~(\ref{e.app.reptation.<u(t)u(t)u(0)u(0)>}),
  (\ref{e.app.reptation.<uuuu>}) and (\ref{e.app.reptation.<uu><uu>}) into
  Eq.~(\ref{e.app.reptation.<p(t)p(t)p(0)p(0)>}), we have
\begin{align}
  \label{e.app.reptation.<p(t)p(t)p(0)p(0)>.f}
  \left\langle p_i(t)p_j(t) p_k(0) p_l(0) \right\rangle
  &=
  \frac {b^4}{9}
  \left[
  (\delta_{ik}\delta_{jl} + \delta_{il}\delta_{jk})
  \int_0^Nds\int_0^Nds' \Psi(s,s';t)
  + \delta_{ij}\delta_{kl} N^2
  \right],
\end{align}
\end{widetext}
where we used $\Psi(s,s;t) = \Psi(s;t)$.

Taking contractions in Eq.~(\ref{e.app.reptation.<p(t)p(t)p(0)p(0)>.f}) between
the first and second indices, and also between the third and fourth indices, we
obtain
\begin{equation}
  \label{e.app.reptation.<p^2(t)p^2(0)}
  \left\langle \bm{p}^2(t) \bm{p}^2(0) \right\rangle
  =
  \frac {2b^4}{3}
  \int_0^Nds\int_0^Nds' \Psi(s,s';t)
  + \left\langle \bm{p}^2 \right\rangle^2,
\end{equation}
where we used $\left\langle \bm{p}^2 \right\rangle = b^2N$. Inserting
Eq.~(\ref{e.app.reptation.<p^2(t)p^2(0)}) into Eq.~(\ref{e.phi1.reptation}), we
obtain \cite{uneyama15}
\begin{equation}
  \label{e.app.reptation.phi1}
  %\left\langle | \bm{p}(t) \cdot \bm{p}(0)|^{2} \right\rangle
  \phi_1(\tau)
  =
  \frac {2}{3N^2}\int_{0}^{N}ds\int_{0}^N ds' \Psi(s,s';t).
\end{equation}
Similarly, taking contractions in
Eq.~(\ref{e.app.reptation.<p(t)p(t)p(0)p(0)>.f}) between the first and fourth
indices, and also between the second and third indices, we obtain
\begin{equation}
  \label{e.app.reptation.<[p(t)p(0)]^2>}
  \left\langle \left[\bm{p}(t) \cdot \bm{p}(0)\right]^2 \right\rangle
  =
  \frac {4b^4}{3}
  \int_0^Nds\int_0^Nds' \Psi(s,s';t)
  +
  \frac {\left\langle \bm{p}^2 \right\rangle^2}{3},
\end{equation}
By inserting Eq.~(\ref{e.app.reptation.<[p(t)p(0)]^2>}) into
Eq.~(\ref{e.phi2.reptation}), $\phi_2(\tau)$ can be expressed as
\begin{equation}
  %\left\langle | \bm{p}(t) \cdot \bm{p}(0)|^{2} \right\rangle
  \label{e.app.reptation.phi2}
  \phi_2(\tau)
  =
  \frac {4}{N^2}\int_{0}^{N}ds\int_{0}^N ds' \Psi(s,s';t).
\end{equation}
By comparing Eqs.~(\ref{e.app.reptation.phi1}) and (\ref{e.app.reptation.phi2}),
we obtain $\phi_2(\tau) = 6\phi_1(\tau)$ [Eq.~(\ref{e.reptation.phi_2=6phi_1})].

\section {Langevin equation of COM motion for rigid rod-like polymer}
\label{s.app.rod}

Here, Eq.~(\ref{e.rodlike-polymer.B(t)}) for the rigid rod-like polymer is
derived.  The overdamped COM motion of the rod is described as follows
\cite{dhont96}:
\begin{equation}
  \label{e.app.rod.dR/dt}
  \frac {d \bm{R}_G(t)}{dt}
  =
  \bm{\Gamma}^{-1}_{f} \cdot \bm{f}(t, \hat{\bm{u}}(t)),
\end{equation}
where $\bm{\Gamma}_f^{-1}$ is the inverse of the friction matrix, namely
\begin{equation}
  \label{e.app.rod.Gamma_f}
  \bm{\Gamma}^{-1}_f
  =
  \frac {1}{\zeta_{\parallel}}
  \hat{\bm{u}}(t)\hat{\bm{u}}(t)
  +
  \frac {1}{\zeta_{\perp}}
  \left[
  \hat{\bm{I}} - \hat{\bm{u}}(t)\hat{\bm{u}}(t)
  \right],  
\end{equation}
and $\zeta_{\parallel}$ and $\zeta_{\perp}$ are the friction coefficients
parallel and perpendicular to the rod's long axis, respectively.

Note that the thermal noise $\bm{f}(t, \hat{\bm{u}}(t))$ depends on the
direction $\hat{\bm{u}}(t)$ of the rod. This noise term can be decomposed as
\begin{equation}
  \label{e.app.rod.f(t,u)}
  \bm{f}(t, \hat{\bm{u}}(t))
  =
  \bm{f}_{\parallel}(t, \hat{\bm{u}}(t))
  +
  \bm{f}_{\perp}(t, \hat{\bm{u}}(t)),
\end{equation}
where $\bm{f}_{\parallel}(t, \hat{\bm{u}}(t))$ and $\bm{f}_{\perp}(t,
\hat{\bm{u}}(t))$ represent equilibrium thermal noise in the parallel and
perpendicular directions of the rod:
\begin{align}
  \label{e.app.rod.f_parallel}
  \bm{f}_{\parallel}(t, \hat{\bm{u}}(t))
  &=
  \left(2\zeta_{\parallel} k_BT\right)^{1/2}
  \hat{\bm{u}}(t)\hat{\bm{u}}(t)
  \cdot
  \bm{\xi}(t),
  \\[0.1cm]
  \label{e.app.rod.f_perp}
  \bm{f}_{\perp}(t, \hat{\bm{u}}(t))
  &=
  \left(2\zeta_{\perp} k_BT\right)^{1/2}
  \left[
  \hat{\bm{I}} - \hat{\bm{u}}(t)\hat{\bm{u}}(t)
  \right]
  \cdot
  \bm{\xi}(t).
\end{align}
Here, $\bm{\xi}(t)$ is the three-dimensional white Gaussian noise defined in
Eq.~(\ref{e.def-lefd.<xi(t)xi(0)>}). Note that $\bm{\xi}(t)$ is independent of
$\hat{\bm{u}}(t)$, in contrast to $\bm{f}(t, \hat{\bm{u}}(t))$ in
Eq.~(\ref{e.app.rod.dR/dt}). Inserting
Eqs.~(\ref{e.app.rod.Gamma_f})--(\ref{e.app.rod.f_perp}) into
Eq.~(\ref{e.app.rod.dR/dt}) and using the Einstein relations $D_{\parallel} =
k_BT / \zeta_{\parallel}$ and $D_{\perp} = k_BT / \zeta_{\perp}$, we have
Eq.~(\ref{e.rodlike-polymer.B(t)}).

%section {bibliography}

%\bibliography{paper}

%merlin.mbs apsrev4-1.bst 2010-07-25 4.21a (PWD, AO, DPC) hacked
%Control: key (0)
%Control: author (8) initials jnrlst
%Control: editor formatted (1) identically to author
%Control: production of article title (-1) disabled
%Control: page (0) single
%Control: year (1) truncated
%Control: production of eprint (0) enabled
%

\end {document}